\documentclass[sigconf, nonacm]{acmart}
\newcommand\vldbdoi{XX.XX/XXX.XX}
\newcommand\vldbpages{XXX-XXX}
\newcommand\vldbvolume{19}
\newcommand\vldbissue{11}
\newcommand\vldbyear{2026}
\newcommand\vldbauthors{\authors}
\newcommand\vldbtitle{\shorttitle} 
\newcommand\vldbavailabilityurl{https://github.com/saltsystemslab/jasper}
\newcommand\vldbpagestyle{empty} 

\usepackage{svg}
\usepackage{tabularx}

\usepackage{graphicx} 
\usepackage[ruled,vlined]{algorithm2e}
\usepackage{amsmath}
\usepackage{hyperref}
\usepackage{cleveref}
\usepackage{svg}

\usepackage[subtle]{savetrees}

\usepackage[textsize=tiny]{todonotes} 
\usepackage[normalem]{ulem} 

\usepackage{graphicx} 
\usepackage{amsmath}
\usepackage{amstext}
\usepackage{amsfonts}
\usepackage{mathrsfs}
\usepackage{xcolor}
\usepackage{tikz}
\usetikzlibrary{shapes}
\usetikzlibrary {shapes.geometric}
\usetikzlibrary{calc}
\usepackage{cleveref}
\usepackage{subcaption}
\usepackage{pgfplots}
\usepackage{pgfplotstable}
\pgfplotsset{compat=1.3}
\usepackage{url}
\usepackage{xspace}
\usepackage{siunitx}
\usepackage{booktabs}
\usepackage{multirow}
\usepackage{comment}
\usepackage{enumitem}
\setitemize{leftmargin=*}
\usetikzlibrary{external}
\usepackage{graphicx}
\usepackage{svg}
\usepackage{balance}
\usepackage{placeins}  
\usepackage{footmisc}
\usepackage[normalem]{ulem}
\usepackage{cancel}
\usepackage{array}

\title{GPU-Accelerated ANNS: Quantized for Speed, Built for Change}

\author{Hunter McCoy}
\affiliation{%
  \institution{Northeastern University}
  \city{Boston}
  \country{United States}
  \postcode{02115}
}
\email{mccoy.hu@northeastern.edu}

\author{Zikun Wang}
\affiliation{%
  \institution{Northeastern University}
  \city{Boston}
  \country{United States}
  \postcode{02115}
}
\email{wang.zikun@northeastern.edu}

\author{Prashant Pandey}
\affiliation{%
  \institution{Northeastern University}
  \city{Boston}
  \country{United States}
  \postcode{02115}
}
\email{p.pandey@northeastern.edu}

\newif\ifshowcomments
\showcommentsfalse  

\newif\iftruncate

\truncatetrue   

\newif\ifappendix
\appendixtrue   

\newcommand{\apxref}[2]{\ifappendix\Cref{#1}\else\cite[#2]{McCoy2026}\fi}

\ifshowcomments
    \newcommand{\prashant}[1]{{\scriptsize \textcolor{red}{Prashant: {#1}}}}
    \newcommand{\hunter}[1]{{\scriptsize \textcolor{blue}{Hunter: {#1}}}}
    \newcommand{\zikun}[1]{{\scriptsize \textcolor{purple}{Zikun: {#1}}}}
\else
    \newcommand{\prashant}[1]{}
    \newcommand{\hunter}[1]{}
    \newcommand{\zikun}[1]{}
\fi

\newcommand{\para}[1]{\smallskip\noindent\textbf{#1.}}

\newcommand{\defn}[1]{{\textit{\textbf{\boldmath #1}}}\xspace}

\newcommand{\sysname}{Jasper\xspace}
\newcommand{\sysnameRQ}{Jasper RaBitQ\xspace}
\newcommand{\RQ}{RaBitQ\xspace}
\newcommand{\cagra}{CAGRA\xspace}
\newcommand{\ganns}{GANNS\xspace}
\newcommand{\parlayann}{ParlayANN\xspace}
\newcommand{\diskann}{DiskANN\xspace}
\newcommand{\bang}{BANG\xspace}

\newcommand{\bigann}{BigANN\xspace}
\newcommand{\deep}{Deep\xspace}
\newcommand{\gist}{Gist\xspace}
\newcommand{\openai}{OpenAI\xspace}
\newcommand{\textToImage}{Text2Image\xspace}

\newcommand{\figsize}{.24}

\newcommand{\fastestBuild}{1,179K\xspace}
\newcommand{\timesFasterThanCagra}{1.84\xspace}
\newcommand{\lowerBangSpeedup}{10\xspace}
\newcommand{\higherBangSpeedup}{74\xspace}
\newcommand{\fastestQuery}{12.5 million\xspace}
\newcommand{\fastestOpenAIQuery}{3.2 million\xspace}
\newcommand{\latencyReduction}{14\%\xspace}

\newcommand{\avgConstruct}{7.0$\times$\xspace}

\newcommand{\prune}{\textsc{robustPrune}\xspace}
\newcommand{\search}{\textsc{beamSearch}\xspace}

\ifshowcomments
    \newcommand{\marginComment}[1]{\marginpar{\sf \scriptsize \textcolor{red}{#1} }}
    \newcommand{\RTwoInline}[2]{{\textcolor{forestGreen}{\marginComment{R2: #1}{#2}}}}
    \newcommand{\RThreeInline}[2]{{\textcolor{blue}{\marginComment{R3: #1}{#2}}}}
    \newcommand{\RFourInline}[2]{{\textcolor{magenta}{\marginComment{R4: #1}{#2}}}}
\else
    \newcommand{\marginComment}[1]{}
    \newcommand{\RTwoInline}[2]{{#2}}
    \newcommand{\RThreeInline}[2]{{#2}}
    \newcommand{\RFourInline}[2]{{{#2}}}
\fi

\begin{document}

\begin{abstract}

Approximate nearest neighbor search (ANNS) is a core problem in machine
learning and information retrieval. GPUs offer a promising path to
high-performance ANNS through massive parallelism and co-location with
downstream applications, but current GPU indices face three limitations:
inability to update without full rebuilds, lack of efficient quantization for
high-dimensional vectors, and poor latency hiding due to data-dependent memory
accesses.

We present \sysname, a GPU-native ANNS system built on the Vamana graph index
that achieves both high query throughput and full updatability via three new
techniques: (1) a batch-parallel construction algorithm enabling lock-free
streaming insertions, (2) a GPU-efficient \RQ implementation that reduces
memory footprint up to 8$\times$ without random access penalties, and (3) an
optimized search kernel with improved compute utilization and latency hiding.

Across five datasets, \sysname achieves up to \timesFasterThanCagra$\times$
higher throughput than \cagra, the current
state-of-the-art GPU index, while providing updatability that \cagra lacks,
constructs indices \avgConstruct faster on average, and delivers
\lowerBangSpeedup–\higherBangSpeedup$\times$ faster queries than \bang, the
previous fastest GPU Vamana implementation.

\end{abstract}

\maketitle

\pagestyle{\vldbpagestyle}
\begingroup\small\noindent\raggedright\textbf{PVLDB Reference Format:}\\
\vldbauthors. \vldbtitle. PVLDB, \vldbvolume(\vldbissue): \vldbpages, \vldbyear.\\
\href{https://doi.org/\vldbdoi}{doi:\vldbdoi}
\endgroup
\begingroup
\renewcommand\thefootnote{}\footnote{\noindent
This work is licensed under the Creative Commons BY-NC-ND 4.0 International License. Visit \url{https://creativecommons.org/licenses/by-nc-nd/4.0/} to view a copy of this license. For any use beyond those covered by this license, obtain permission by emailing \href{mailto:info@vldb.org}{info@vldb.org}. Copyright is held by the owner/author(s). Publication rights licensed to the VLDB Endowment. \\
\raggedright Proceedings of the VLDB Endowment, Vol. \vldbvolume, No. \vldbissue\ %
ISSN 2150-8097. \\
\href{https://doi.org/\vldbdoi}{doi:\vldbdoi} \\
}\addtocounter{footnote}{-1}\endgroup

\ifdefempty{\vldbavailabilityurl}{}{
\vspace{.3cm}
\begingroup\small\noindent\raggedright\textbf{PVLDB Artifact Availability:}\\
The source code, data, and/or other artifacts have been made available at \url{\vldbavailabilityurl}.
\endgroup
}

\pgfplotsset{
  recallPlot/.style={
    small,
    width = 1.2\columnwidth,
    height = .7\columnwidth,
    table/col sep=comma,
    xlabel near ticks,
    x label style={at={(0.5,-0.2)},font=\small},
    y label style={at={(-0.2,0.5)},font=\small},
    ytick style={draw=none},
    ytick scale label code/.code = {},
    yticklabel style={ /pgf/number format/fixed },
    ymajorgrids,
    minor tick num=1,
    minor grid style={draw=gray!25},
    xmax=1.02,
    xmin=-0.02
  },
}

\pgfplotsset{
  ablationPlot/.style={
    small,
    width = 1.1\columnwidth,
    height = 0.9\columnwidth,
    table/col sep=comma,
    xlabel near ticks,
    x label style={at={(0.5,-0.2)},font=\small},
    y label style={at={(-0.2,0.5)},font=\small},
    ytick style={draw=none},
    ytick scale label code/.code = {},
    yticklabel style={ /pgf/number format/fixed },
    ymajorgrids,
    minor tick num=1,
    minor grid style={draw=gray!25},
    xmax=1.02,
    xmin=-0.02
  },
}

\pgfplotsset{
  rooflinePlot/.style={
    small,
    width = 1.1\columnwidth,
    height = 1.0\columnwidth,
    table/col sep=space,
    xlabel near ticks,
    x label style={at={(0.5,-0.3)},font=\small},
    y label style={at={(-0.2,0.5)},font=\small},
    ytick style={draw=none},
    ytick scale label code/.code = {},
    yticklabel style={ /pgf/number format/fixed },
    ymajorgrids,
    minor tick num=1,
    minor grid style={draw=gray!25},
    xmin=0.01, xmax=1000,
    ymin=0.1, ymax=100,
  },
}

\pgfplotsset{
  quantizationPlot/.style={
    small,
    width = 1.1\columnwidth,
    height = .7\columnwidth,
    ylabel={Throughput},
    table/col sep=comma,
    xlabel near ticks,
    x label style={at={(0.5,-0.2)},font=\small},
    y label style={at={(-0.15,0.5)},font=\small},
    ytick style={draw=none},
    ytick scale label code/.code = {},
    yticklabel style={ /pgf/number format/fixed },
    ymajorgrids,
    minor tick num=1,
    minor grid style={draw=gray!25},
    xmax=1.05,
  },
}

\pgfplotsset{
  insertPlot/.style={
    small,
    width = 1.1\columnwidth,
    height = .7\columnwidth,
    ylabel={Throughput (inserts/s)},
    table/col sep=comma,
    xlabel near ticks,
    x label style={at={(0.5,-0.12)},font=\small},
    y label style={at={(-0.15,0.5)},font=\small},
    ytick style={draw=none},
    ytick scale label code/.code = {},
    yticklabel style={ /pgf/number format/fixed },
    ymajorgrids,
    minor tick num=1,
    minor grid style={draw=gray!25},
    xmax=1.05,
  },
}

\pgfplotsset{
  incrementalBuildPlot/.style={
    small,
    width = .85\columnwidth,
    height = .4\columnwidth,
    ylabel={Throughput (inserts/s)},
    table/col sep=comma,
    xlabel near ticks,
    x label style={at={(0.5,-0.12)},font=\small},
    y label style={at={(-0.15,0.5)},font=\small},
    ytick style={draw=none},
    ytick scale label code/.code = {},
    yticklabel style={ /pgf/number format/fixed },
    ymajorgrids,
    minor tick num=1,
    minor grid style={draw=gray!25},
  },
}

\pgfplotsset{
  incrementalCompPlot/.style={
    small,
    width = .85\columnwidth,
    height = .4\columnwidth,
    ylabel={Update Time (Seconds)},
    table/col sep=comma,
    xlabel near ticks,
    x label style={at={(0.5,-0.12)},font=\small},
    y label style={at={(-0.15,0.5)},font=\small},
    ytick style={draw=none},
    ytick scale label code/.code = {},
    yticklabel style={ /pgf/number format/fixed },
    ymajorgrids,
    minor tick num=1,
    minor grid style={draw=gray!25},
  },
}

\pgfplotsset{
    ablationBaseStyle/.style={color=gray!60, mark=*, mark size=1.5pt, thick, dashed},
    ablationBlock64Style/.style={color=gray!80!blue, mark=square*, mark size=1.5pt, thick},
    ablationNoHashStyle/.style={color=blue!50, mark=triangle*, mark size=1.5pt, thick},
    ablationTile4Style/.style={color=blue!70, mark=diamond*, mark size=1.5pt, thick},
    ablationChunkedStyle/.style={color=blue!90, mark=pentagon*, mark size=1.5pt, thick},
    ablationRabitqStyle/.style={color=red!80!orange, mark=*, mark size=2pt, very thick},
  }


\definecolor{rose}{HTML}{B8336A}

\definecolor{turq}{HTML}{48D1CC}

\definecolor{medGreen}{HTML}{027148}

\definecolor{lightGreen}{HTML}{D3F7AD}

\definecolor{khaki}{HTML}{6A5F31}

\definecolor{skysurge}{HTML}{5BC0EB}

\definecolor{bananacream}{HTML}{FDE74C}

\definecolor{magenta}{HTML}{822E81}

\definecolor{rosewood}{HTML}{AA6373}

\definecolor{mintGreen}{HTML}{98FF98}

\definecolor{gold}{HTML}{FFB703}

\definecolor{strongred}{HTML}{D62828}

\definecolor{forestGreen}{HTML}{228C22}

\definecolor{strongblue}{HTML}{FF7F0F}

\definecolor{weakpurple}{HTML}{38B2AC}

\definecolor{weakblue}{HTML}{3A86FF}

\definecolor{weakgray}{HTML}{285D34}

\definecolor{weakcyan}{HTML}{1982C4}

\pgfplotsset{
  jasperStyle/.style = {color = strongred, mark = square*, mark size=1.5pt, thick},
  jasperRqStyle/.style = {color = strongblue, mark = square*, mark size=1.5pt, thick},
  jasperHostStyle/.style = {color = red, mark = *, mark size=1.5pt, , dash pattern = on 1.5pt off 1pt, thick},
  cagraStyle/.style = {color = weakpurple, mark = *, mark size=1.5pt, thick},
  parlayStyle/.style = {color = weakblue, mark = triangle*, mark size=1.5pt, thick},
  bangStyle/.style = {color=weakgray, mark = diamond*, mark size=1.5pt, thick},
  gannsStyle/.style = {color=weakcyan, mark = pentagon*, mark size=1.5pt, thick},
  warpcoreStyle/.style = {color=forestGreen, mark = square*},
  p2MetaStyle/.style = {color=black, mark = diamond*},  cuckooStyle/.style = {color=turq,mark=square*},
  doubleMetaStyle/.style = {color=brown, mark = triangle*},
  bchtStyle/.style = {color=cyan, mark = square*},
  p2bhtStyle/.style = {color=lightGreen, mark = square*},
  slabStyle/.style = {color=gold, mark = square*},
  p2ExtStyleFill/.style = {fill = red},
  p2IntStyleFill/.style = {fill=cyan},
  doubleStyleFill/.style = {fill=orange},
  ihtStyleFill/.style = {fill=magenta},
  ihtMetaStyleFill/.style = {fill=medGreen},
  chainingStyleFill/.style = {fill=violet},
  warpcoreStyleFill/.style = {fill=forestGreen},
  p2MetaStyleFill/.style = {fill=black},
  doubleMetaStyleFill/.style = {fill=brown},
  cuckooStyleFill/.style = {fill=turq},
  jasperStyleFill/.style = {fill=strongred},
  cagraStyleFill/.style = {fill = weakpurple},
  jasperThickStyle/.style = {color = strongred, mark = *, mark size=1.5pt, thick},
  jasperRqThickStyle/.style = {color = strongblue, mark = *, mark size=1.5pt, thick},
  cagraThickStyle/.style = {color = weakpurple, mark = *, mark size=1.5pt, thick},
}

\section{Introduction}

Nearest neighbor search is a fundamental problem in machine learning and
information retrieval. The problem involves finding the $k$ closest points to a
query in a high-dimensional space. In this paradigm, complex data such as
images, documents, or user profiles are embedded as vectors in a \emph{metric
space} (such as Euclidean, Manhattan, Hamming) that enables similarity to be
measured through geometric distance. Applications span recommendation
systems~\cite{Roy2022, Singh2021}, image
retrieval~\cite{Rahman2024OptimizingDI, Belahyane2020}, anomaly
detection~\cite{Schubert2015, Okkels2024, Gu2019}, and increasingly,
retrieval-augmented generation (RAG) for large language models~\cite{Quinn2025,
Wang2024, Wang2025}, where relevant context must be retrieved from massive
corpora in real time.

Traditional indexing structures such as $k-d$ trees~\cite{Bentley1975} and quad
trees~\cite{Finkel1974} provide efficient exact nearest neighbor search in low
dimensions, but scale poorly as dimensionality increases. \iftruncate The
\emph{curse of dimensionality}~\cite{Minsky2017} causes distances between
points to concentrate, with the ratio between nearest and farthest neighbors
approaching one. This phenomenon undermines the spatial pruning heuristics
that traditional indices rely upon, causing query complexity to degrade
toward exhaustive linear scans. Modern embedding models routinely produce
vectors with hundreds to thousands of dimensions, limiting the practicality
of exact methods. \fi

Approximate nearest neighbor search (ANNS) addresses this challenge by relaxing
the exactness constraint.  \RFourInline{D1}{Instead of guaranteeing the true
nearest neighbors, ANNS returns approximate results whose accuracy can be
tuned against query cost. This trade-off makes it possible to scale
efficiently to high-dimensional datasets.} The three dominant paradigms for
ANNS are: inverted file indices (IVF)~\cite{Babenko2015, Xia2013, Sivic2003,
Jgou2011}, locality-sensitive hashing (LSH)~\cite{Indyk1998, Datar2004,
Motwani2008}, and graph-based indices~\cite{Malkov2020, Malkov2014, Dong2011,
Subramanya2019,Gollapudi2023, Manohar2024, ootomo2024, FAISS}.
\RFourInline{D1}{To improve performance further, these techniques can be
combined with vector quantization, which reduces vector size while preserving
relative distances.}

Graph-based indices achieve the best recall-throughput trade-offs among these
paradigms. Recent empirical studies have consistently demonstrated this
advantage~\cite{Subramanya2019}. Graph-based indices construct a
\emph{navigable search graph}~\cite{Malkov2014} where vertices correspond to
vectors and edges encode neighborhood relationships. Nearest neighbor queries
proceed via greedy traversal also known as \emph{beam search}: starting from a
designated entry point, the search iteratively expands the closest unvisited
vertex until convergence, maintaining a frontier of the best (top $k$)
candidates encountered.

\para{Modern applications require updatable ANNS indexes} Real-world ANNS
applications operate at massive scale and require continuous updates due to
evolving datasets. Web-scale retrieval systems index billions of vectors:
Facebook's similarity search infrastructure handles over 200 billion
embeddings~\cite{FAISS}, while Microsoft's Bing uses DiskANN to search over a
billion document vectors~\cite{adams2025distributedann}. These systems must
support both millions of queries per second and continuously index updates as
new content arrives~\cite{Singh2021}. Music streaming, product recommendation,
and RAG systems depend on scalable and updatable similarity
search~\cite{Singh2021a}.



\para{The case for GPU acceleration} GPUs offer massive parallelism that
promises to accelerate ANNS to achieve high throughput and low latency. Both
index construction and search are well-suited to leverage this parallelism
because the core operations (pair-wise distance computations) are independent
and can execute concurrently across candidates. 
As dataset size and vector dimensionality grow, so does the available parallel
work that GPUs can exploit.

%
Furthermore, GPUs are already widely deployed in settings where ANNS is most
needed. Modern machine learning infrastructure runs on GPU clusters. Using GPUs
for ANNS allows organizations to leverage existing hardware rather than
provisioning separate CPU-based retrieval infrastructure. This co-location also
reduces data movement: embeddings generated on GPU can be indexed and queried
without costly transfers to CPU memory.

\para{Challenges in exploiting GPU parallelism for ANNS} Modern GPUs offer
thousands of execution units and memory bandwidth exceeding 2
TB/s~\cite{a100specs}, yet achieving peak performance for ANNS indexing and
queries remains difficult. We identify three key challenges in exploiting the
massive GPU parallelism. 

1) \textit{ANNS workloads need updatability.} Most current GPU indices require
a rebuild from scratch in order to process new updates. 
However, real-world applications such as anomaly detection~\cite{Gu2019} and
RAG~\cite{Quinn2025, Wang2024, Wang2025} require the ability to modify the
index to incorporate new information in search results.

2) \textit{Greedy graph traversal creates irregular memory access patterns.}
Unlike dense matrix operations where access patterns are known ahead of time,
graph search follows data-dependent paths that vary per query. When different
threads in a warp access non-contiguous memory locations, the GPU must issue
multiple memory transactions, wasting memory bandwidth. This problem is made
worse by the pointer-chasing nature of graph traversal, where each hop depends
on the previous result.

3) \emph{ANNS workloads are memory-bound rather than compute-bound.} Each
vector dimension is read once, used for a single multiply-add in the distance
computation, then discarded. With no data reuse, performance is limited by how
fast data can be fetched from memory rather than how fast arithmetic can be
performed. For a 128-dimensional float32 vector, the GPU performs 256
floating-point operations but must load 512 bytes of data, a ratio far below
what modern GPUs can sustain at peak compute utilization. In our analysis
in~\Cref{eval:roofline}, we find that the state-of-the-art ANNS search kernels
are very close to the peak memory capacity. \RFourInline{D1}{To further improve performance, we need to reduce the data transfer and avoid incurring random memory accesses. Efficient vector quantization techniques can help us achieve both targets.}


\para{Limitations of existing GPU ANNS systems} 
Existing GPU ANNS systems either lack updatability or suffer from poor
performance. \cagra~\cite{ootomo2024}, the current state-of-the-art GPU ANNS
system, achieves high query throughput by employing a carefully optimized graph
structure derived from NN-Descent. However, \cagra's graph construction
algorithm is batch-oriented, and the system is not designed for incremental
updates. In production settings where data arrives continuously, the inability
to update the index without full reconstruction is a critical limitation. 

BANG~\cite{Venkatasubba2025} is based on an updatable graph structure but
suffers from severe performance penalties as it employs product quantization
(PQ)~\cite{Jgou2011}~\footnote{PQ is a quantization technique that reduces
  vector size by decomposing the original vector into a Cartesian product of
independently quantized low-dimensional subspaces.} for compressing the
vectors. \bang extends the Vamana ANNS index~\cite{Subramanya2019}; this index
structure enables updates (although \bang currently does not implement index
updates). However, its reliance on Product Quantization (PQ) limits peak
performance.
In~\Cref{sec::quant}, we discuss in detail the limitations of product
quantization on GPUs.


Other GPU ANNS systems face their own constraints. GANNS~\cite{Sun2024}
implements HNSW but struggles with the algorithm's sequential construction
dependencies. SONG~\cite{Zhao2020} and GTS~\cite{Zhu2024} target different
index structures (fixed-rank graphs and trees, respectively) with different
trade-off profiles. None of these systems simultaneously achieve
state-of-the-art query performance, efficient construction, and support for
incremental updates.

\para{This paper} We present \sysname, a GPU-native ANNS system that addresses
the above challenges to achieve high query throughput while supporting dynamic
index updates. \sysname builds on graph-ANNS techniques from
DiskANN~\cite{Subramanya2019} and ParlayANN~\cite{Manohar2024}. It identifies a
set of design principles for minimizing data movement and maximizing
utilization on irregular, memory-bound GPU workloads. Each principle is counter
to established techniques in CPU ANNS systems. We make the following
contributions:

\begin{itemize}

  \item \textbf{Batch-parallel GPU construction with support for incremental
    updates.} We devise a batch-parallel construction
    algorithm for GPU execution. Rather than using fine-grained locks that
    create serialization bottlenecks at high-degree vertices, our approach
    performs beam searches independently, accumulates candidate edges, sorts
    them by target vertex, and applies pruning in parallel without needing
    synchronization. \RTwoInline{D1}{The same algorithm drives both initial
    bulk construction and incremental inserts after the index is
    live, making \sysname the first GPU ANNS system to support 
    updates without a full rebuild.}

  \item \textbf{GPU-efficient quantization.} Quantization for GPUs is an
    open problem. Product Quantization (PQ)~\cite{Jgou2011}, the default
    scheme in prior GPU ANNS systems, incurs heavy read amplification and
    random memory accesses on every codebook lookup. We present the first GPU
    implementation of RaBitQ~\cite{Gao2024}, a quantization scheme that
    compresses vectors through randomized rotation and scalar quantization for
    up to 8$\times$ memory reduction; its sequential loads and simple
    arithmetic increase arithmetic intensity.

  \item \textbf{Optimized search kernel.} 
    Beam search conventionally uses a
    hash table to avoid revisiting vertices, but on a GPU this costs 512B of
    shared memory per query and caps occupancy; removing it costs only a few
    redundant distance computations while freeing shared memory to schedule
    more concurrent queries per block. We find that for smaller vectors block
    size and occupancy dominate performance, with peak throughput at 32 threads
    per block (a single warp) and 100\% occupancy, maximizing memory-level
    parallelism and hiding pointer-chasing latency (defined
    in~\Cref{subsec:memory}).


  \item \textbf{Optimized memory access patterns.} We introduce a tile-based
    loading scheme that issues blocked 16-byte loads across the threads in a
    warp simultaneously. Naive implementations either serialize loads through a
    single thread or issue element-wise loads across the warp (defined
    in~\Cref{subsec:memory}). Our approach partitions vectors into 16-byte
    chunks that are processed in parallel.

  \item \textbf{Systematic performance analysis.} We conduct a detailed
    empirical study of GPU ANNS optimization strategies. We characterize the
    trade-offs between per-query parallelism and memory-level parallelism,
    finding that optimal block sizes differ significantly across datasets.
    \iftruncate
    We
    perform roofline analysis~\cite{Williams2009} on the state-of-the-art
    search kernels and find that modern ANNS kernels are memory bound and
    execute at close to optimal throughput.
  \fi

\end{itemize}

\para{Results}
Our evaluation on five standard ANNS benchmarks demonstrates:

\begin{itemize}

  \item \sysname achieves up to \timesFasterThanCagra$\times$ higher query
    throughput than \cagra, the state-of-the-art GPU index, while providing
    updatability that \cagra lacks.

  \item Compared to \bang, the previous state-of-the-art GPU Vamana
    implementation, \sysname delivers
    \lowerBangSpeedup–\higherBangSpeedup$\times$ higher throughput.

  \item On an A100, \sysname achieves peak throughput exceeding \fastestQuery queries/sec
    on the \bigann dataset and over \fastestOpenAIQuery queries/sec on
    1,536-dimensional \openai embeddings with RaBitQ quantization.

  \item \sysname's has the fastest incremental and batch GPU construction:
    \sysname has a peak insertion throughput of \fastestBuild insertions per
    second and is \avgConstruct faster on average than \cagra for index
    construction across five datasets.

  \item Our coalesced vector loading scheme improves memory access performance by up
    to \latencyReduction at low beam widths, with no throughput penalty at high
    beam widths.

\end{itemize}

\para{Broader implications for ANNS on GPUs}
\RFourInline{D3}{The underlying techniques proposed in our work address fundamental challenges in GPU-accelerated ANNS workloads.}
Our roofline analysis confirms that state-of-the-art kernels for \RFourInline{D3}{ANNS} already operate near the memory bandwidth ceiling, suggesting that future gains must come primarily from reducing data movement rather than algorithmic improvements alone.

\section{Preliminary}

This section describes the GPU execution and memory models relevant to building
high-throughput ANNS indexes. We first explain the GPU execution model and how
work is scheduled on the GPU. We then explain the GPU memory hierarchy and how
it affects achieving peak performance. We then examine how these hardware
characteristics affect existing GPU ANNS systems and introduce the roofline
model~\cite{Williams2009} as a framework for analyzing their performance.

\begin{figure}[t]
    \centering
    \begin{subfigure}{0.49\linewidth}
        \centering
        \includegraphics[width=\linewidth]{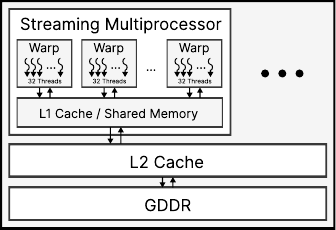}
        \caption{GPU Memory Hierarchy}
        \label{fig:gpu-memory}
    \end{subfigure}
    \hfill
    \begin{subfigure}{0.49\linewidth}
        \centering
        \includegraphics[width=\linewidth]{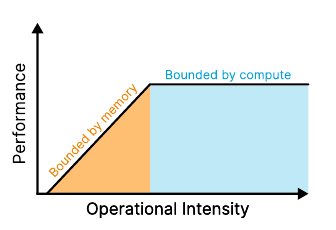}
        \caption{Roofline model}
        \label{fig:roofline-model}
    \end{subfigure}
    \caption{GPU memory hierarchy and roofline model.}
    \vspace{-1em}
\end{figure}

\subsection{GPU execution model} 

A diagram of the GPU architecture is shown in~\Cref{fig:gpu-memory}. A GPU
consists of multiple \defn{streaming multiprocessors (SMs)}, each capable of
hosting up to 1024 concurrent threads. 
Tasks on a GPU are executed in a chunk of work called a \textbf{kernel}. Each
kernel is a single-instruction, multiple-thread (SIMT) function that is
executed in parallel by every thread. GPU kernels are executed as collections
of \emph{thread blocks} on SMs, each containing up to 1024 threads. Within a
block, threads are subdivided into \emph{warps} (groups of 32 threads) that
share execution resources and issue memory operations cooperatively. As a
result, performance is maximized when threads within a warp follow identical
control flow and access contiguous or nearby memory locations. Warps can be
further subdivided into \emph{tiles} that divides the resources within a warp
for multiple jobs.

ANNS workloads expose fine-grained parallelism during nearest neighbor search
as pairwise distances can be computed independently in parallel. Each vector is
examined by a \emph{tile}, allowing multiple candidate vectors to be accessed
and evaluated in parallel. 
\iftruncate
Tiles are densely packed into warps: larger tiles increase per-probe
parallelism by examining more individual vectors faster, while smaller tiles
improve latency hiding by enabling additional concurrent memory requests within
a warp.
\fi

The optimal block and tile configuration depends on the characteristics of the
ANNS index, including memory layout, probe fan-out, and candidate filtering
costs.

\subsection{GPU memory model} \label{subsec:memory} 

\iftruncate
The design of concurrent GPU data structures is fundamentally constrained by
the underlying memory consistency model, which governs how memory operations
from different threads become visible to one another. In contrast to CPUs that
are sequentially consistent, modern GPUs implement a weak memory model based on
Relaxed Memory Ordering (RMO)~\cite{Alglave2015}. Under RMO, reads and writes
may be freely reordered unless explicitly constrained, and a read in one thread
that logically follows a write in program order from a different thread is not
guaranteed to observe that write without additional synchronization. GPUs
prioritize throughput over latency. With thousands of concurrent threads on
GPUs, enforcing strong ordering requires expensive coherence traffic.
\fi





Each streaming multiprocessor contains an L1 cache shared among all threads in the SM. This shared L1 cache is crucial for achieving high performance in GPU
applications. A section of this cache can be designated as \textbf{shared
memory}. Unlike a regular cache, this shared memory is directly addressable and
acts as both a scratchpad and a means of efficiently passing data between
threads in an SM. The amount of shared memory utilized affects the potential
\textbf{occupancy}, \RTwoInline{D2}{the ratio of active warps to the maximum
number of warps an SM can schedule.} Every block in a kernel requests the same
amount of shared memory; if the SM cannot fit enough blocks to keep its warp
slots full, occupancy drops and fewer warps run in parallel, leaving threads
idle.


To achieve high performance on GPUs, we must design GPU kernels for ANNS index
construction and querying that efficiently utilize shared memory to maximize
occupancy. Specifically, we aim to keep all frequently used data in shared
memory and carefully select data structures that fit within the available
resources without reducing maximum occupancy.


\subsection{Roofline model}

The roofline model~\cite{Williams2009} is a simple, quantitative performance
model that bounds the attainable performance of an application on a given
architecture using two hardware limits: peak compute throughput and peak memory
bandwidth. Performance is expressed as a function of \emph{arithmetic
intensity}, defined as the ratio of useful operations to bytes transferred from
main memory. For low arithmetic intensity kernels, performance is limited by
memory bandwidth and grows linearly with arithmetic intensity; for high
intensity kernels, performance saturates at the hardware’s peak compute rate.
Graphically, these bounds form a ``roofline'' (see~\Cref{fig:roofline-model}):
a sloped bandwidth roof intersecting a flat compute roof, making it easy to see
whether a workload is memory-bound or compute-bound and how close it is to
hardware limits.


The roofline model is widely used to guide performance optimization and
architectural analysis as it abstracts complex microarchitectural details into
a small set of measurable parameters. Optimizations that increase data reuse
(e.g., blocking, caching, fusion) move kernels rightward by increasing
arithmetic intensity, while vectorization, parallelism, and instruction-level
efficiency move kernels upward towards the compute roof. 
\iftruncate
Extensions of the model incorporate multiple memory levels, ceilings for
specific bottlenecks, and heterogeneous devices, but the core insight remains:
achievable performance is fundamentally constrained by the balance between
computation and data movement. This framing has proven effective for reasoning
about performance portability across CPUs, GPUs, and accelerators, and for
communicating optimization priorities in both systems and data-intensive
workloads. \emph{We use the roofline model to evaluate both the achieved and
  potential performance of ANNS indices on GPUs.}
\fi

\section{Vamana Index}
\label{sec::diskann}


In this section, we cover the Vamana algorithm graph structure, construction
and query algorithm, and the batch-parallel construction algorithm. 

Our system
\sysname employs the Vamana graph~\cite{Subramanya2019}, a directed proximity
graph designed to support efficient greedy search with bounded out-degree, to
build the approximate nearest neighbor search (ANNS) index. We choose the
Vamana graph as it supports both a fast and accurate greedy search and an
efficient parallel construction algorithm introduced in
ParlayANN~\cite{Manohar2024}. 
We describe how we develop GPU-accelerated Vamana index along with several GPU
optimizations in the next section.



\subsection{Graph definition} 
The Vamana index is a directed graph $G = (V, E)$ constructed over a set of
data vectors $V = \{x_1, \dots, x_N\}$, where $N$ is the total number of
vectors. Each node maintains an adjacency list of at most $R$ outgoing edges. A
distinguished entry point $s \in V$ is used to initialize both graph
construction and query traversal.

The edge set $E$ is selected to ensure that the graph remains sparse while
preserving navigability. For a given vertex in the graph and query vector,
there is a high probability that one of the outgoing edges of the vertex leads
to a vertex that is closer to the query. This property means that greedy search
will converge to a set of points close to a target query and is critical for
enabling low-latency approximate search.

\subsection{Vamana construction algorithm}

The original Vamana construction algorithm starts with all vertices contained
within a graph with a random set of edges with maximum outgoing degree $R$.
Vamana graph construction proceeds incrementally to refine the graph one vertex
at a time. Refinement of verticies proceeds in two phases. First, Vamana uses
\search to identify candidate neighbors in the existing graph for the new
vertex. Second, Vamana applies a pruning procedure, \prune, to reattach the
vertex to the graph while enforcing a strict bound on the out-degree.



\begin{algorithm}
\DontPrintSemicolon
\caption{BeamSearch (DiskANN)}
\label{alg:beamsearch}
\KwIn{Graph $G$, query vector $q$, beam width $K$, start vertex $m$ (medoid)}
\KwOut{Frontier $F$ of $K$ nearest neighbors, visited set (hash table) $V$}
$F \gets \{m\}$ (priority queue by distance) \;
$V \gets \emptyset$ \;
\While{there exists an unvisited vertex $u \in F$}{
    Mark $u$ visited: $V \gets V \cup \{u\}$ \;
    \For{each neighbor $v$ of $u$}{
        \If{$v \notin V$}{
            Compute distance $d(q,v)$ \;
            Insert $v$ into $F$ with priority $d(q,v)$ \;
        }
    }
    \If{$|F| > K$}{
        Remove furthest elements from $F$ until $|F|=K$ \;
    }
}
\Return $(F, V)$
\end{algorithm}

\para{Beam search} Given a vector $x$ to be inserted, the construction algorithm starts with a search for the approximate nearest neighbors using \search (\Cref{alg:beamsearch}). \search starts at a designated start point, typically chosen as the medoid, and maintains two sets: (i) a \emph{frontier} which stores the $K$ closest candidate vertices discovered so far, and (ii) a \emph{visited list} which records all vertices whose outgoing edges have been explored. In each iteration, the algorithm \emph{expands} the closest frontier vertex not yet visited, loading the candidates from its outgoing edges, computing their distance to $x$, and inserting them into the frontier.



If the size of the frontier exceeds the beam width $K$, the furthest candidates
are discarded. This expansion process continues until all vertices in the
frontier have been visited. 
The procedure returns both the final frontier and the full list of visited vertices.


\para{Edge attachment}
Once \search completes, the new vertex $x$ is attached to the graph using the visited list. Bidirectional edges are added
between $x$ and each vertex in the visited list, temporarily allowing the
out-degree of affected vertices to exceed the maximum degree bound $R$.

\para{Robust prune}
To restore the degree bound, Vamana applies the \prune procedure to any vertex
whose outgoing edge list exceeds $R$ (maximum out degree). 
Given a vertex $p$, \prune sorts the outgoing neighbors by increasing
distance from $p$.
The algorithm iteratively selects the closest remaining neighbor $p^*$ and
adds it to the pruned adjacency list. For each remaining candidate neighbor
$p'$, \prune compares the distance $d(p^*, p')$ to the distance
$d(p, p')$. If
\[
\alpha \cdot d(p^*, p') \le d(p, p'),
\]
then $p'$ is removed from consideration. This process repeats until either the
pruned adjacency list reaches size $R$ or no candidates remain.

\subsection{ParlayANN}
\label{sec::parlay}

\parlayann~\cite{Manohar2024} is an extension of the Vamana construction
algorithm that addresses a fundamental scalability bottleneck in graph construction. The original \diskann~\cite{Subramanya2019} inserts vertices in parallel using fine-grained locks, leading to significant contention on high-centrality vertices such as the
\emph{medoid}.
\parlayann eliminates this bottleneck by replacing lock-based updates with a
lock-free, batch-parallel construction strategy. The core idea is to decouple
graph traversal from graph mutation, allowing expensive updates to be applied
in parallel without contention.



\parlayann restructures graph construction into two
distinct, parallelizable phases. In the first phase, \search
(\Cref{alg:beamsearch}) is executed independently in parallel for each new vertex.

Each search operates on a read-only snapshot of the current graph and adjacency lists are
not modified during this phase.
Instead of applying updates immediately, all candidate edges produced by the
batch are written to a temporary edge collection. This collection is then
\emph{semisorted}, grouping edges by their target vertex. Semisort guarantees that all candidate edges incident to
a given vertex are co-located, enabling subsequent processing to be performed
independently per vertex.

In the second phase, \prune  (\apxref{alg:robustprune}{Appendix~A}) is applied in
a batch-parallel design. Each vertex that has received new candidate edges is
assigned a unique thread, which merges the proposed edges with the existing
adjacency list and applies \prune. Because all edges incident to a vertex are processed by a
single thread, this phase requires no locks or cross-thread synchronization.

This lock-free design removes contention around centrally located vertices and
allows \parlayann to efficiently exploit multicore parallelism.


\begin{algorithm}
\DontPrintSemicolon
\caption{\sysname BeamSearch \hspace{0.4em}\textnormal{\small (refer to \Cref{alg:beamsearch})}}
\label{alg:jasperbeam}
\KwIn{Graph $G$, query $q$, beam width $K$, start vertex $m$}
\KwOut{Frontier $F$ of $K$ nearest neighbors (each entry carries a visited flag)}

$F \gets \{m\}$ \tcp*{priority queue by $d(q,\cdot)$; visited flag per entry}
\While{some entry in $F$ is unvisited}{
    $u \gets$ closest unvisited entry in $F$; mark $u$ visited\;
    \tcp{Gather: all neighbors of $u$ (no visited filter --- duplicates allowed)}
    $C \gets N(u)$\;
    \tcp{Distance: parallel tile-based chunked loads compute $d(q,v)$}
    \ForEach{$v \in C$ \textnormal{in parallel across the block's tiles}}{
        compute $d(q, v)$\;
    }
    \tcp{Sort + merge: full sort over the union}
    $F \gets \textsc{Sort}(F \cup C)$ by $d(q, \cdot)$\;
    \tcp{Prune: clip to top-$K$ (visited flags carry over)}
    truncate $F$ to $K$\;
}
\Return $F$
\end{algorithm}



\subsection{Why is the Vamana appropriate for GPUs?} \label{sec::why-vamana}


Compared to many alternative ANN indices, the Vamana graph offers a high degree
of parallelism across both index construction and query execution, enabling
efficient use of modern GPUs. Its construction is inherently parallel: vertices
are processed independently, each following a predictable sequence of steps
that minimizes thread divergence. Distance computations and graph traversal
both expose fine-grained parallelism. Vamana also supports efficient
incremental updates. Each insertion performs a bounded beam search over a small
subset of vertices and adds at most 2R edges, keeping overhead and space usage predictable. By
comparison, update procedures for methods such as NN-Descent can be
substantially more expensive, as multiple refinement rounds are required and
any visited node may become a neighbor. Prior work shows that Vamana matches or
exceeds alternative graph indices in search quality across diverse datasets,
making it a strong foundation for high-performance, updatable
ANNS~\cite{Subramanya2019}.

%
%

\section{\sysname: A GPU native ANNS system}
\label{subsec:gpu-optimizations}

\RFourInline{W1}{\sysname's design targets four GPU characteristics to achieve
  high performance. The
  central challenge in GPU-based ANNS is sustaining high hardware (compute and
  memory) utilization for an irregular, data-dependent, and memory-intensive
  workload. Existing systems trade off compute efficiency for memory bandwidth,
  or vice versa. \sysname achieves both simultaneously by re-architecting
  greedy beam search and its associated data structures around GPU execution
  constraints. The four characteristics addressed by \sysname are: }

  \begin{itemize}
    \item \RFourInline{W1}{\emph{Shared memory:} per-query state (frontier,
      candidate buffer, distances) lives in shared memory, and \sysname
      aggressively optimizes to reduce shared memory usage and increase occupancy. The
      removal of the visited hash table employed in \parlayann and \cagra helps more queries
      fit per SM.}
    \item \RFourInline{W1}{\emph{Warp divergence:} the irregular phases of beam
      search are confined to a single warp to minimize control-flow divergence.}
    \item \RFourInline{W1}{\emph{Register pressure:} distance computation is
      tiled into 16-byte chunks (\Cref{fig::load_comparison}), bounding the
      per-thread accumulator state to the register budget needed for peak
      occupancy at 32 threads/block.}
    \item \RFourInline{W1}{\emph{Memory access pattern:} our GPU \RQ implementation
      uses sequential global-memory loads, avoiding the random
      codebook lookups and read amplification incurred by PQ
      (\Cref{sec::quant}). For high-dimensional datasets, we
      widen the block to trade memory-level parallelism for per-query compute.}
  \end{itemize}





\begin{figure}[t]
    \centering
    \includegraphics[width=0.95\linewidth]{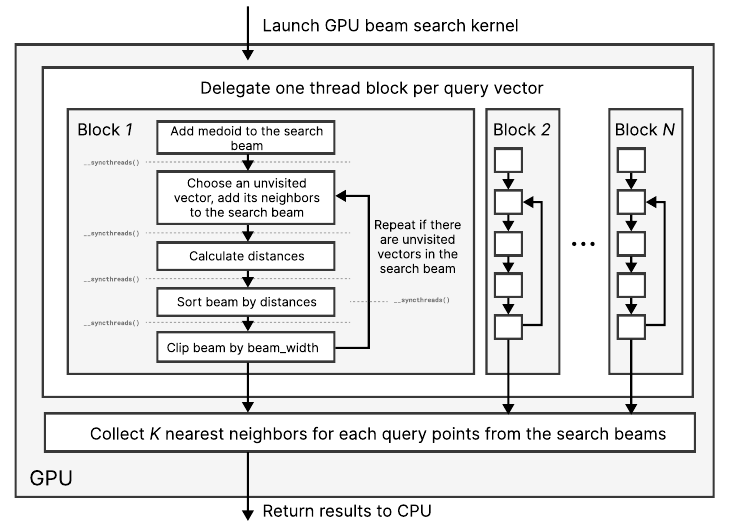}
    \caption{ANNS query pipeline in \sysname.}
    \label{fig:query-gpu}
\end{figure}

\subsection{Block-per-query execution model}

A central design decision in \sysname is to assign one CUDA thread block to
each query or inserted vector. This block-per-query model allows threads to
cooperatively execute search while keeping all per-query state (frontier,
candidates, distances) in shared memory, reducing global memory traffic and
enabling data reuse. Computing distances between vectors is embarrassingly
parallel and occupies the full block, while irregular phases such as candidate
expansion and graph traversal are partially serialized to reduce control-flow
divergence.

The full query pipeline can be seen in \Cref{fig:query-gpu}. 
The size of the thread block controls a trade-off between per-query compute
throughput and the number of memory requests issued. Larger blocks increase
arithmetic throughput, instruction-level parallelism, and shared-memory
capacity per query, while smaller blocks allow more concurrent queries in each
SM (streaming multiprocessor), allowing for higher memory utilization. The
optimal trade-off choice depends on the compute and memory balance of the
distance kernel. For low-dimensional datasets such as \bigann, where
performance is memory-bound, smaller blocks maximize throughput. For
high-dimensional datasets such as \gist, where distance computation dominates
(17.6$\times$ higher cost than \bigann), larger blocks that maximize per-query
parallelism achieve higher throughput. \RTwoInline{D3}{Block size is chosen as
32 for vectors with up to 128 dimensions, and as 64 threads per block for all
larger vectors. For the small vectors, we additionally double the block size at
beam widths of 512 and above as it provides a 38\% performance boost for these
large beam widths.} In summary, \emph{smaller vectors favor
higher query concurrency; larger vectors require greater per-query compute
resources.} 

\subsection{Divergence from \parlayann construction} \label{subsec:divergence}

Our version of \search diverges from \parlayann's \search due to the
difference between CPU and GPU parallelism. We are limited by the computing
model and resources when running \search on GPU. 
The \search algorithm in \parlayann is serial for each query, as there are many
more queries than CPU threads. On the GPU, we have an order of magnitude more
threads. To fully saturate the GPU compute and achieve high performance, we must
parallelize the internal components of \search. 



\para{Different hash table choices} A major source of divergence arises from
\parlayann’s use of a lossy hash table during beam search. The original
algorithm employs a fixed-size (512-slot) hash table with direct replacement to
track previously visited vertices and prune duplicates. On CPUs, this structure
is accessed sequentially within a single thread, yielding deterministic
behavior. On GPUs, preserving this access pattern would require serializing
hash table operations across cooperating threads, severely limiting throughput.
Parallelizing these operations improves performance but introduces
non-determinism. We find that this hash table does not
significantly affect accuracy or search throughput on the GPU,
\RTwoInline{D4}{and remove the hash table in all \sysname configurations,
instead opting to use a visited bit for each item in the frontier.}
\RTwoInline{D5}{Removing the
hash table reduces shared memory pressure and allows \sysname to
schedule more queries per block, as shared memory is the limiting occupancy
factor when running with 32 threads per block.}

\para{Different merge order} \sysname employs a different candidate merge order
during beam search. \parlayann implements a deferred-merge optimization, which
postpones sorting and merging when candidate lists are smaller than a fixed
fraction of the beam width. The size of the candidate set depends on the
results of the hash table lookup: with non-deterministic hash tables, we cannot
guarantee this mechanism produces the same candidate list order each time.
Empirically, we find that removing the hash table and deferred merging
materially affects neither index quality or query accuracy. Consequently,
\sysname removes both features. This simplifies control flow, preserves
determinism, and eliminates a major source of thread divergence.


\subsection{Diverging from \parlayann's search}
\RTwoInline{W1}{\sysname inherits the two-phase batch-parallel structure above
  but tunes the inner \search and the batch-pruning prologue for GPU execution.
\Cref{alg:jasperbeam} shows \sysname's \search; it can be read against
\Cref{alg:beamsearch} to see the changes directly. The visited-tracking hash
table that \parlayann uses to filter neighbors at gather time is removed
entirely: each frontier entry instead carries a visited flag, gather pulls in
\emph{all} neighbors of $u$ without filtering, and any already-visited
duplicates are absorbed by the subsequent sort. This trades a small number of
recomputed distances for $\sim$512 bytes of shared memory per query (the
fixed-size hash table in \parlayann), which raises the number of concurrent
queries per SM. \parlayann's deferred-merge optimization is also removed:
every round, \sysname \emph{gathers}, \emph{computes distances} in parallel
across the block's tiles, \emph{sorts and merges} into $F$, and \emph{prunes}
$F$ back to $K$.
A third change touches the batch pruning prologue (\apxref{alg:parlayann}{Appendix~A},
Step 3): the per-batch \textsc{Semisort} is replaced by a full Thrust
\textsc{Sort} keyed on \emph{(target vertex, distance)}, which both groups edges
per vertex and pre-orders each group for \prune. This single sort yields
better load balance on the GPU than a semisort followed by per-vertex thread
sorts.
}

\subsection{Vector load optimization}

\begin{figure}[!t]
    \centering
    \includegraphics[width=\linewidth]{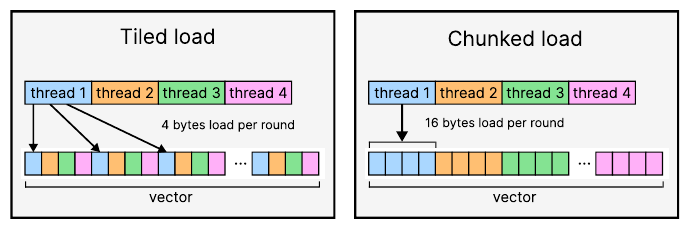}
    \caption{Comparison of the two different load strategies tested in
    \sysname: tiled loads load one object per thread in the tile at a time,
  whereas the chunked load strategy loads a fixed amount of data (16 bytes) per
thread and has each thread work on all objects loaded in that chunk.}
    \label{fig::load_comparison}
    \vspace{-0.5em}
\end{figure}
Efficient distance computation is limited not only by arithmetic throughput but
also by the memory load latency. By default, when a warp accesses
a large object, the compiler emits a sequence of 16-byte loads issued by a
single thread, which limits latency hiding. Element-wise loads distribute
requests across threads but reduce effective bandwidth, as each element is no
more than 4 bytes large.

We adopt a \emph{chunked loading scheme} that partitions each vector into 16-byte
segments and assigns one segment to each thread. Threads load query and
candidate chunks in parallel, compute partial distances locally, and iterate
until the full vector is processed. A block-wide reduction then produces the
final distance. This approach issues many coalesced 16-byte loads
simultaneously to achieve near-peak memory bandwidth.

\Cref{fig::load_comparison} shows the difference between tiled and chunked loading. Chunked loading improves performance for
latency-sensitive workloads: for small beam widths, where search is memory
latency bound, it yields up to a 14\% speedup. For large beam widths, where
compute throughput dominates, optimized loads match baseline performance
without loss.


\begin{figure}[t]
    \centering
    \includegraphics[width=0.95\linewidth]{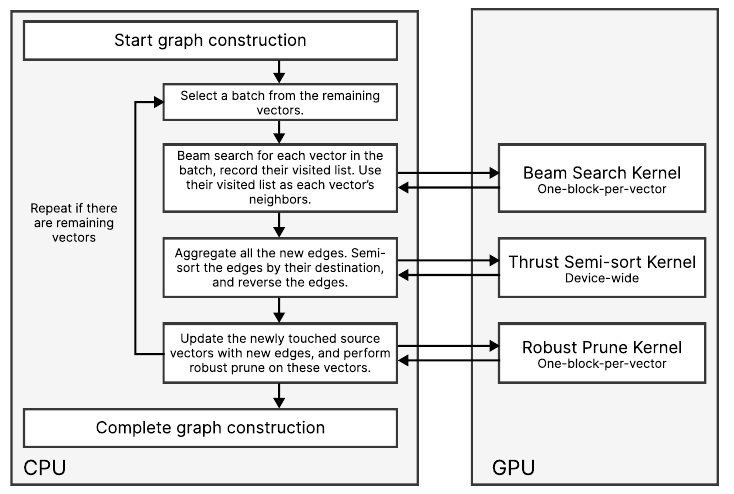}
    \caption{Batch graph construction pipeline in \sysname.}
    \label{fig:batch-construction}
\end{figure}

\subsection{Batch construction on GPUs}

To support high-throughput index construction, \sysname adopts the bulk
construction algorithm \RThreeInline{W3}{as this matches the structure of the original algorithm and is highly amenable to the GPU launch model}. \parlayann relies on fork--join parallelism (via
ParlayLib~\cite{Blelloch2020}) and a semisort to group edges before insertion,
enabling lock-free parallel updates. Fork--join dynamically decomposes large
tasks (e.g., sorting and grouping edges) into smaller subtasks that are
scheduled across CPU threads.

\parlayann's execution model does not map directly to GPUs: kernel launches
require a fixed thread configuration, dynamic task spawning is not supported
within a kernel, and excessive kernel launches incur substantial overhead. We
therefore fuse \emph{search}, \emph{grouping}, and \emph{pruning} into a small
number of self-contained GPU kernels with fixed parallel structure. 
This fused pipeline is illustrated in~\Cref{fig:batch-construction}.
Although this removes some scheduling flexibility, it preserves the scalability of batch-parallel
search and pruning while conforming to the GPU execution model and
minimizing launch overhead.
\RTwoInline{W3,D1}{We employ the batch-parallel design and fuse the build steps
into a fixed kernel pipeline to achieve high performance. Therefore,
insertions and queries are run in phases: a batch of inserts must complete
before queries resume, and vice versa. Inserts are still incremental in the
sense that each batch extends a live index without rebuilding it.} 


\para{GPU-efficient pruning}
We implement a hybrid pruning strategy for \prune. Outgoing edges produced by
beam search are accumulated together with existing edges that require pruning
and are stored in a unified buffer. This buffer is fully sorted by vertex ID and
distance using the CUDA Thrust library. While \parlayann performs a
semisort followed by per-vertex distance sorting using a single thread, we find that on
GPUs a full parallel sort provides better load balance. as individual threads
are relatively weak and uneven work distribution is costly.

After sorting, each vertex’s edge list is pruned by assigning an entire SM
(1024 threads) to the operation. \prune is dominated by distance
computations and therefore benefits from high intra-vertex parallelism.

\subsection{Deletion}

\RThreeInline{W2}{Along with batch insertions, \sysname implements batch deletions using the algorithm from FreshDiskANN~\cite{Singh2021}. In this algorithm, vertices are first logically marked as deleted. This is a lightweight, in-place operation that prevents vertices from appearing in the output but does not remove them from the graph. A bulk remove operation \texttt{compact} is then called once a large percentage of the graph has been deleted (10\%). This culls nodes from the graph without sacrificing recall by using a bulk prune strategy: each vertex that pointed to a deleted vertex is pointed to every directed neighbor of that vertex, and robust prune is called to ensure only high-quality neighbors are kept. This preserves recall and allows the vertex to be entirely removed from the graph. Performance of these phases can be seen in \Cref{subsec::deletion}.}

\subsection{Scalability with batch size for updates}


Index construction in \sysname is performed entirely on the GPU:
\RTwoInline{W2}{both the dataset vectors and the evolving graph reside in
device memory}, eliminating PCIe transfers and using the remaining memory as
workspace for batch processing. The achievable batch size is therefore
constrained by the available device memory after storing the index and directly
determines the degree of parallelism and throughput of the construction
pipeline.

On an NVIDIA A100 with 80 GB of memory, the BigANN-100M dataset and its graph
occupy 38.5 GB, leaving 41.5 GB for construction buffers; this supports an
optimal batch size of approximately 1M vectors. 


\RTwoInline{W2}{
 For Deep 100M datset, during Jasper construction we spend 38.5GB to store the
 raw vectors, 25.6GB to store the graph index, and an additional 8GB for the
 construction workspace. In total, \sysname uses 72GB of device memory during
 construction, which is very close to the 80GB limit by Nvidia A100 GPU. In
 contrast, \cagra's NN-descent construction requires around 150GB of device
 memory for the same dataset, almost double the available memory. In our
 experiments, \cagra failed to allocate sufficient memory for NN-descent and had
 to fall back to the IVF method to build the index trading off construction
 performance and accuracy.
}

\RTwoInline{W2}{
 Overall, \sysname's construction method scales better compared to NN-descent
 when the GPU memory is the bottleneck for large datasets. We calculate the
 full memory usage breakdown in \apxref{app:memory}{Appendix~D}. We also shown in
 \apxref{app:h200}{Appendix~C.2} we can support 200M vector dataset using GPU with larger
 memory.
}

\section{Vector quantization}
\label{sec::quant}

Modern embedding models produce vectors spanning hundreds to thousands of
dimensions, with per-vector sizes ranging from 96 bytes to over 6000 bytes as
shown in \bigann benchmarks~\cite{Simhadri2022}. ANNS is
memory-bound: each vector element is read once, used for a single arithmetic
operation, and discarded. Reducing vector size directly improves query
throughput.

%

To reduce per-vector storage and I/O costs while preserving search accuracy, we
consider \emph{quantization}, which compresses vectors while approximately
preserving distances. The most widely used approach for ANNS is \emph{product
quantization} (PQ)~\cite{Jgou2011}. PQ divides each vector into $K$ disjoint
subvectors and quantizes each independently using a small codebook of 256
centroids. Each subvector is then encoded as a single byte indicating its
nearest centroid, compressing a high-dimensional vector to just $K$ bytes.
Distance computation uses a precomputed \emph{lookup table} storing
centroid-to-centroid distances for each subspace. This results in significant
compression with modest accuracy loss, as error arises only from the
approximation of each subvector by its nearest centroid.


Although PQ compresses vectors effectively, its \emph{lookup table} access
pattern results in lower throughput on GPUs. GPU memory is organized into
32-byte \textit{sectors} that are loaded atomically—reading a single 4-byte
distance entry wastes the remaining 28 bytes, causing $8\times$ read
amplification. A straightforward solution is to place the lookup table in
shared memory, but even a modest configuration ($K=32$) requires
$32\times256\times256\times4$ bytes (8 MB), far exceeding shared memory
capacity. We implemented PQ in \sysname but found that throughput was 
strictly worse than unquantized search for all configurations tested. Similar
conclusions regarding the PQ performance on GPU has been presented in prior
research~\cite{ootomo2024}.

\subsection{RaBitQ quantization}



\begin{figure}[t]
    \centering
    \includegraphics[width=\columnwidth]{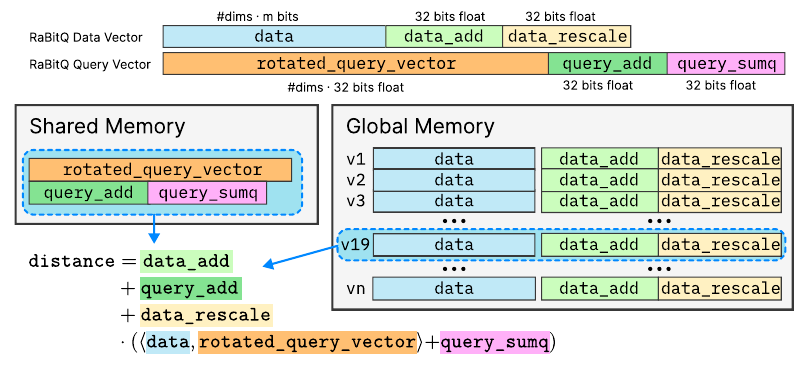}
    \caption{During L2 distance calculation, RaBitQ loads query vector and its
    metadata from shared memory, and it loads quantized data vector and its
  metadata from global memory.}
    \label{fig:rabitq_load}
\end{figure}

To effectively use quantization on the GPU, we must avoid random memory
accesses during distance computation. Previous GPU implementations treat
quantization as a trade-off: smaller vectors and lower memory I/O at the cost
of lower throughput~\cite{Venkatasubba2025, ootomo2024}. We show this trade-off
is not fundamental. \sysname employs RaBitQ~\cite{Gao2024}, a quantization
method that is well suited for GPUs. RaBitQ requires only sequential memory
access, simple arithmetic, and requires no lookup tables.


RaBitQ estimates the distance using the inner product between the quantized
data vector with the query vector. 
We define the following variables: $m$ is the number of quantized
bits per dimension, $c$ is the centroid vector, $v$ is the original data vector,
$q$ is the original query vector, $o$ is the rotated and normalized data vector 
$v-c$, $\bar{o}$ is the quantized data vector, and $\Delta_x$ is the rescaling 
factor of $\bar{o}$ from $o$.

During data quantization, RaBitQ compresses the
data vectors by first normalizing them and applying a random rotation matrix.
With high dimensional vectors, any rotated dimensions are likely tightly
clustered around 0: according to the Johnson-Lindenstrauss
lemma~\cite{Johnson1984}, the value of any coordinate is no further than
$\frac{2}{\sqrt{D}}$ away from 0 with high probability. This allows RaBitQ to
treat each dimension as an unbiased distribution and enables simple scalar
quantization. RaBitQ encodes each dimension with $m$ bits along with two
floating point metadata: \texttt{data\_add} and \texttt{data\_rescale}
, where they correspond to  $||v-c||^2 + 2\cdot ||v-c||^2 \cdot {\langle c, \bar{o} 
\rangle \over \langle o, \bar{o} \rangle}$ and $-2{\Delta_x \over \langle o, \bar{o}
\rangle}$. These precomputed values save computation during distance estimation.
In total, the size of the quantized vector is $\text{\#dims} \cdot m\,\text{
bits} + 2\cdot\mathrm{sizeof}(\text{float})$.

During querying, RaBitQ applies the same rotation matrix to the query vector,
and RaBitQ calculates each query vector's metadata: \texttt{query\_add} and
\texttt{query\_sumq}, where they correspond to $||q-c||^2$ and $\sum (q-c) 
\cdot {2^{m-1}-1\over 2}$. \Cref{fig:rabitq_load} shows how RaBitQ estimates
distances solely based on quantized vectors and rotated query vectors. The
RaBitQ estimator reduces the L2 distance calculation to estimating the inner
product between the quantized data vector (\texttt{data}) and rotated query
vector (\texttt{rotated\_query\_vector}). This process estimates the vector
distance without branching or randomized memory accesses while most of the
values are pre-computed, which makes it highly suitable for GPU execution.
RaBitQ's sequential memory access during distance calculation also allows us to
utilize the same loading technique for exact vector loading
in~\apxref{sec::microbenchmark}{Appendix~E}.

\sysname integrates GPU-accelerated RaBitQ. RaBitQ achieves $2\times$
compression for 8-bit vectors and up to $8\times$ for 32-bit vectors. On
datasets with large dimensions, our results demonstrate that \sysname with
RaBitQ performs significantly better than our exact distance calculation
version. In~\apxref{sec::microbenchmark}{Appendix~E}, we perform a detailed analysis between
RaBitQ's performance and PQ's performance.

\section{Evaluation} 


In this section, we evaluate the performance of \sysname
against four state-of-the-art ANNS indices: three GPU-based and one
CPU-based index. We evaluate the construction, query, and update performance
across five real-world datasets varying in size and number of dimensions. Wex
further evaluate the effectiveness of \sysname in saturating GPU resources using
the roofline model~\cite{Williams2009}. Finally, we perform a series of
micro-benchmarks to evaluate the impact of various GPU-specific optimizations
introduced in \sysname and discuss how these are broadly applicable for
optimizing other GPU-accelerated systems.

\subsection{Experimental setup}

We evaluate all systems on along three dimensions:
\begin{itemize}
    \item \textbf{Construction throughput:} measured as the total time required
      to build the index used in subsequent experiments.
    \item \textbf{Query throughput:} reported as the number of ANN queries
      completed per second.
    \item \textbf{Recall:} which measures the accuracy of the Top-\textit{K}
      results and is defined as $\texttt{Recall@}K = \frac{\texttt{\# of exact
      Top-\textit{K} results returned}}{K}$. We report recall at 1@1, 10@10,
      50@50, and 100@100.
\end{itemize}

\para{Hardware} All CPU experiments are conducted on an Intel(R) Xeon(R) Gold 5218 CPU with 32 physical cores (64 hardware threads) at 2.30 GHz. GPU
experiments are run on an NVIDIA A100 (SM80) with 80 GB of HBM memory using
CUDA 12.9.

\setlength{\tabcolsep}{2pt}
\begin{table}
\begin{center}
\resizebox{\columnwidth}{!}{
\begin{tabular}{|c|c c c c|} 
 \toprule
 Name & Data Type & Dimensions & Distance Type & Size \\  
 \midrule
 \bigann & \texttt{uint8} & 128 & Euclidean & 100M\\ 
 \deep & \texttt{float32} & 96 & Euclidean & 100M\\
 \gist & \texttt{float32} & 960 & Euclidean & 1M \\
 OpenAI-ArXiv & \texttt{float32} & 1536 & Euclidean & 2.3M \\
 Yandex Text-to-Image & \texttt{float32} & 200 & Inner product & 10M\\
 \bottomrule
\end{tabular}
}
\end{center}
\caption{Details of datasets employed in our evaluation.}
\label{fig::datasets}
\vspace{-2em}
\end{table}

\para{Datasets} For evaluation, we select the following datasets taken from the
Big\_ANN\_Benchmarks~\cite{Simhadri2022} repository, summarized
in~\Cref{fig::datasets}.
\begin{itemize}
  \item \bigann: 100 million 128-dimensional \texttt{uint8} vectors with Euclidean
      distance from the SIFT collection. There are 10K vectors in the
      test set.
    \item \deep/Yandex: 100 million PCA-reduced, normalized GoogLeNet
      embeddings (96-dimensional  \texttt{float32}). Euclidean distance and 10K
      queries.
    \item \gist: 1 million 960-dimensional \texttt{float32} vectors with euclidean
      distance and 1K query vectors in the test set.
    \item OpenAI-Arxiv: 2.3 million 1536-dimensional \texttt{float32} embeddings of
      Arxiv papers. Query set is 10K vectors.
    \item \textToImage: 10 million 200-dimensional \texttt{float32} image embeddings
      from the Se-ResNext-101 model. Distance is maximum inner product search
      (MIPS) and there are 10K queries in the test set.
\end{itemize}

\para{ANNS indices} We compare Jasper to the following ANNS indices:
\begin{itemize}

    \item \textbf{\cagra}~\cite{ootomo2024}: The state-of-the-art in GPU index
      construction. \cagra constructs a graph index using
      NN-Descent~\cite{Wang2021}.
      \iftruncate
      NN-Descent iteratively improves graph
      connectivity by expanding each node's neighborhood through 2-hop
      neighbors. The resulting graph is well-connected but too dense for GPU
      memory. \cagra addresses this with a three-phase pruning algorithm: edges
      are sorted by length, redundant edges with good 2-hop alternatives are
      removed, and finally the graph is merged with its reverse and pruned
      again to produce the final index.
      \fi
    
    \item \textbf{\bang}~\cite{Venkatasubba2025}: \bang extends the query
      execution of DiskANN~\cite{Subramanya2019} to GPUs. \bang does not
      perform construction and requires a preconstructed Vamama graph from
      DiskANN.  
      \iftruncate
      It uses product quantization to reduce the size of the
      vectors. The quantized vectors are stored in the GPU memory while the
      graph and full vector data are stored in DRAM.
      The entirety of a
      streaming multiprocessor (SM) is used for each query. One thread is used
      for each distance calculation, as distances can be calculated using only
      PQ table lookups. To process queries more efficiently, Bloom filters are
      used to represent the visited list, lowering the memory requirements for
      searching.
      \fi
    
    \item \textbf{\ganns}: \ganns~\cite{Yu2022} is an implementation of the
      HNSW graph algorithm optimized for GPUs. 
      \iftruncate
      To optimize for GPU execution,
      \ganns uses lazy data structures - rather than maintain a priority queue,
      \ganns maintains two lists (frontier and visited) and only occasionally
      updates the lists to mark vertices as visited. They avoid the use of a
      hash table for marking visited vertices during queries, and use multiple
      thread blocks per query. 
      \fi
      The construction algorithm for HNSW graphs is
      difficult to parallelize, so \ganns constructs many small graphs in
      parallel and then sequentially merges them. \ganns does not support
      dataset with 100 million vectors.
    
    \item \textbf{\parlayann}: \parlayann is the state of the art CPU ANNS system.
      \iftruncate
      It implements Vamana for 
      construction and implements the batch-parallel construction algorithm.
      This algorithm is discussed in more detail in \Cref{sec::parlay}.
      \parlayann has the ability to perform \textit{scalar} quantization in
      which floats are translated directly into smaller integer types by evenly
      bucketing the real number line into buckets of a size $\Delta$ based on
      the integer size.
      \fi
\end{itemize}

For all indices the graph size is held constant at R=64, so each vertex has at
most 64 outgoing edges.

\begin{table}
\begin{center}
\resizebox{\columnwidth}{!}{
\begin{tabular}{|c|c c c c|c|} 
 \toprule
 \multirow{2}{*}{Dataset} & \multirow{2}{*}{\sysname} & \multirow{2}{*}{\cagra} & \multirow{2}{*}{\parlayann} & \multirow{2}{*}{\ganns} & \sysname Speedup \\
 & & & & & over \cagra \\
 \midrule
 \bigann & \textbf{327.7} & 1323.3 & 1418.2 & - & 4.04$\times$ \\
 \deep & \textbf{296.2} & 825.6 & 4336.9 & - & 2.79$\times$ \\
 \gist & \textbf{12.2} & 14.1 & 210.6 & 151.9 & 1.16$\times$ \\
 \openai & \textbf{89.0} & 127.8 & 1379.4 & 899.2 & 1.44$\times$ \\
 \textToImage & \textbf{24.3}& 619.2 & 184.0 & 1017.4 & 25.48$\times$ \\
 \bottomrule
\end{tabular}
}
\end{center}
\caption{Index construction time in seconds for various ANNS indexes. \parlayann is constructed on CPU. \ganns could not finish construction on \bigann and \deep 100M datasets.}
\label{tab::construction}
\end{table}


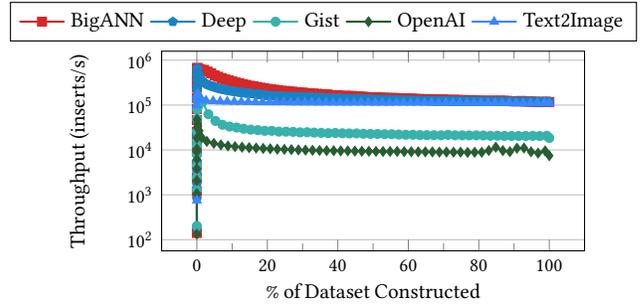
\begin{figure}
\centering
\ref{incrementalLegend}
\begin{tikzpicture}
  \centering
  \begin{axis}[
    incrementalBuildPlot,
    legend to name=incrementalLegend,
    legend columns = 6,
    xlabel={\% of Dataset Constructed},
    legend style={font=\small},
    ymode=log,
    every axis plot/.append style={each nth point=2, filter discard warning=false},
    ]
    
    \addplot[jasperStyle] table [col sep=comma, x=n_points, y=throughput] {data/a100/incremental_build/bigann.txt};

    \addplot[gannsStyle] table [col sep=comma, x=n_points, y=throughput] {data/a100/incremental_build/deep.txt};

    \addplot[cagraStyle] table [col sep=comma, x=n_points, y=throughput] {data/a100/incremental_build/gist.txt};

    \addplot[bangStyle] table [col sep=comma, x=n_points, y=throughput] {data/a100/incremental_build/openai.txt};

    \addplot[parlayStyle] table [col sep=comma, x=n_points, y=throughput] {data/a100/incremental_build/t2i.txt};

 
    \legend{\bigann, \deep, \gist, OpenAI, \textToImage}
    
  \end{axis}
  \end{tikzpicture}
  \caption{\sysname incremental construction throughput. X-axis is the \% of the dataset constructed so far, with each data point being the throughput for one batch. Y-axis is throughput on a log-scale, higher is better.}
\label{fig:incremental_jasper}
\vspace{-1em}
\end{figure}

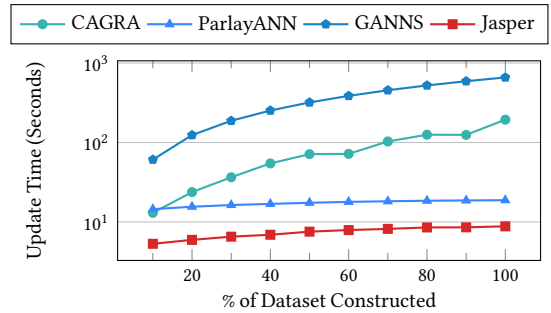
\begin{figure}
\centering
\ref{incrementalCompLegend}
\begin{tikzpicture}
  \centering
  \begin{axis}[
    incrementalCompPlot,
    legend to name=incrementalCompLegend,
    legend columns = 6,
    xlabel={\% of Dataset Constructed},
    legend style={font=\small},
    ymode=log,
    ]

    \addplot[cagraStyle] table [col sep=space, x=n_points, y=runtime_seconds] {data/a100/incremental_vs_full_construction/cagra_bigann.txt};

    \addplot[parlayStyle] table [col sep=space, x=n_points, y=runtime_seconds] {data/a100/incremental_vs_full_construction/parlay_bigann.txt};

    \addplot[gannsStyle] table [col sep=space, x=n_points, y=runtime_seconds] {data/a100/incremental_vs_full_construction/ganns_bigann.txt};

    \addplot[jasperStyle] table [col sep=space, x=n_points, y=runtime_seconds] {data/a100/incremental_vs_full_construction/jasper_bigann.txt};

 
    \legend{\cagra, \parlayann, GANNS, Jasper}
    
  \end{axis}
  \end{tikzpicture}
  \caption{Incremental vs non-incremental construction time for \bigann
  10M. \sysname and \parlayann support batch incremental updates. \cagra and
\ganns do not support updates and need to be built from scratch at each step. X-axis is percentage of dataset constructed and each data point is a further $\frac{1}{10}$th of dataset constructed.
Y-axis is on log-scale, lower is better.}
  \label{fig:incremental_comparison}
  \vspace{-1em}
\end{figure}

\subsection{Index construction}
Here we evaluate the index construction performance of \sysname against other
ANNS indexes. 

\para{Bulk construction} In this benchmark, datasets are constructed in one
shot. Results are shown in~\Cref{tab::construction}. We measure the overall
construction throughput as the ratio of the number of vectors indexed and total
construction time. \sysname achieves the highest construction throughput on all
datasets with peak throughput of \fastestBuild inserts/secs on
\bigann. \parlayann while running on CPU (64 threads) is between $4.3-17.3\times$ slower than \sysname. \ganns is another GPU-based ANNS index but is between $10.1-41.8\times$ slower than \sysname. Additionally, \ganns does not support construction on 100M datasets as it is constrained due to \texttt{int32} datatype being employed to index into the vector array. 100M vectors with 128/96 dimensions require 64-bit datatype for indexing.

    





\begin{figure*}[t]
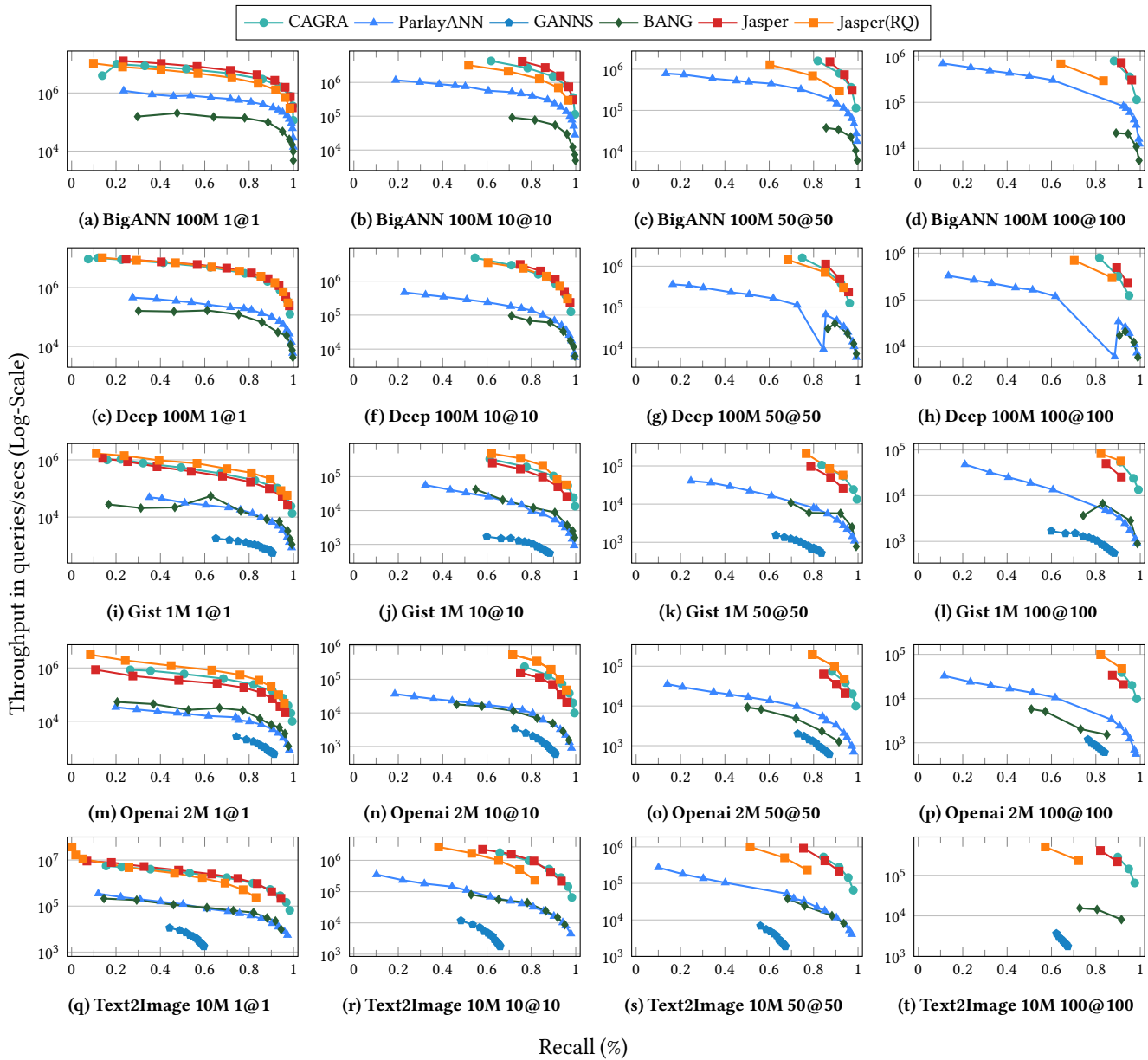

\centering
\ref{queryLegend}

\setlength{\tabcolsep}{0pt}

\begin{tabular}{c c}
\makebox[10pt][c]{\rotatebox[origin=c]{90}{\large Throughput in queries/secs (Log-Scale)}} &
\begin{tabular}{c c c c}

\input{figures/a100-bigann-k1} &
\input{figures/a100-bigann-k10} &
\input{figures/a100-bigann-k50} &
\input{figures/a100-bigann-k100} \\

\input{figures/a100-deep-k1} &
\input{figures/a100-deep-k10} &
\input{figures/a100-deep-k50} &
\input{figures/a100-deep-k100} \\

\input{figures/a100-gist-k1} &
\input{figures/a100-gist-k10} &
\input{figures/a100-gist-k50} &
\input{figures/a100-gist-k100} \\

\input{figures/a100-openai-k1} &
\input{figures/a100-openai-k10} &
\input{figures/a100-openai-k50} &
\input{figures/a100-openai-k100} \\

\input{figures/a100-t2i-k1} &
\input{figures/a100-t2i-k10} &
\input{figures/a100-t2i-k50} &
\input{figures/a100-t2i-k100}

\end{tabular}
\end{tabular}

\vspace{0.5em}
\makebox[\textwidth][c]{\large Recall (\%)}

\caption{Query recall/throughput curves for various ANNS indexes across five datasets on Nvidia A100 GPU. We report recall at 1@1, 10@10, 50@50, and 100@100. Y-axis is on log-scale.
X-axis is recall. Higher and more to the right is better.}
\label{fig::a100_query}
\vspace{-0.5em}
\end{figure*}

We sweep the insertion batch size from 0.5\% to 32\% of the
dataset on \bigann 10M and \gist 1M. Insertion throughput
rises monotonically with batch size and then saturates as the GPU is fully
utilized: on \bigann 10M, throughput grows from $\sim$233K inserts/sec at
0.5\% to $\sim$753K at 16\%, after which 32\% gives no further gain; on \gist
1M it scales smoothly from 52K to 198K going from 0.5--32\%.
Recall is essentially unaffected: across
all beam widths the per-row variation in recall as batch size grows is below
two percentage points (e.g.\ beam-128 recall on \gist drops from 0.854 at
0.5\% to 0.820 at 32\%), confirming that batch size is a throughput knob rather
than a quality knob.

\para{Incremental construction}
\Cref{fig:incremental_jasper} shows how \sysname's construction throughput
changes as the index grows. We build the index incrementally, inserting vectors
in batches of 2\% of the total dataset. Across all datasets, throughput
decreases as the index grows, but the slowdown is sub-linear in index
size. For example, on \deep, \sysname inserts 814K vectors/secs when the index
is at 5\% capacity (5M vectors). When the index reaches 99\% capacity (99
million vectors), throughput drops to 247K vectors/secs. This is roughly a
3.2$\times$ slowdown despite a 20$\times$ increase in index size, demonstrating
that \sysname maintains efficient insertion performance even as the index grows
large.

\Cref{fig:incremental_comparison} illustrates why incremental construction
matters. We measure the time required to add a 10\% slice of new vectors to an
already-constructed index, simulating a common production scenario where new
data arrives after the initial index is built. \sysname and \parlayann, which
support incremental updates, complete this operation an order of magnitude
faster than \cagra and \ganns. The difference arises because \cagra and \ganns
lack incremental update support: each batch of new vectors requires rebuilding
the entire index from scratch.

%


\subsection{Queries}
Here we evaluate the approximate nearest neighbor search (ANNS) performance of
\sysname against other ANNS indexes. \Cref{fig::a100_query} shows the
recall/throughput curves for all indexes across the five datasets on the A100.


\para{Low-dimensional vector datasets: \bigann and \deep} 
For the low-dimensional vector datasets, \sysname is the fastest index for all
recall values supported, with peak throughput of 12.5 million queries / sec at
.23 recall. On \bigann, it is followed by \sysnameRQ which has the second
highest throughput. \sysname is the fastest ANNS index up to .9943 recall as it
does not achieve an accuracy higher than that on \bigann.
For \deep, \cagra is the second fastest index for all recall values tested.
\cagra achieves a higher maximum accuracy on these datasets, with peak accuracy
of .9999 for \bigann 1@1 and .983 for \deep 1@1. 

\sysname is the fastest for the low-dimensional vector datasets due to its
optimized vector loads. For small beam sizes, latency hiding is not as
effective as there is relatively little work performed per vector. This means
that reducing latency of loads improves the performance of these kernels, as
GPU compute spends less time waiting. On these low-dimensional datasets
\sysnameRQ is slightly slower than exact distance computation due to its lossy
representation and additional arithmetic overhead, but it reduces device memory
usage, allowing more vectors to reside on the GPU in memory-constrained
settings.


\para{High-dimensional vector datasets: \gist and \openai}
On the high-dimensional vector datasets \gist and \openai, distance
calculations dominate the run time. To measure the cost of operations, we add
cycle counters to every operation in the search kernel. When running \sysname
exact on both \bigann and \gist, we see a $17.8\times$ increase in the cycles
taken for distance calculations in \gist, with distance calculations becoming
the most expensive operation in the entire search.

The large vector size of these datasets benefits quantization as it can reduce
the compute and memory requirements of distance calculations. On these
datasets, \sysnameRQ is the fastest algorithm for all sizes it supports, with
peak throughput over 3.2 million queries/secs and recall up to .96 for
50@50. While \sysname's exact distance computation is slower than \cagra on
these datasets, \sysnameRQ closes this gap and surpasses \cagra at the same
recalls despite its lossy representation; on \openai at 90\% recall (10@10) it
is 1.6$\times$ faster than exact \sysname. After \sysnameRQ, \sysname has the best throughput-recall trade of curve below .95 recall, at which point \cagra takes over. These systems are followed in order of pareto-optimality by \parlayann and \bang, and finally \ganns.


\RFourInline{W2}{For 100@100 recall on these large vector datasets, \cagra is
able to achieve higher throughput and recall than either \sysname variant: on
\gist, \sysnameRQ stays Pareto-optimal only up to $\sim$0.95 recall, above
which \cagra dominates the throughput--recall curve. Both \gist and \openai exhibit the
same crossover. This is the regime where the lossy RaBitQ representation hits
its accuracy ceiling, and workloads that require very high recall on
high-dimensional vectors are better served by \cagra's exact-distance index.}


\para{\textToImage} \textToImage uses maximal inner
product (MIPS) MIPS for distance instead of Euclidean. This makes construction
and querying this dataset difficult as MIPS does not preserve the triangle
inequality. Having a triangle inequality is necessary for the space to be
metric and is crucial in constructing good search graphs using the Vamana
indexing method. Without this property, graph construction can have poor
results as there is no correlation between the neighbors of a given vector: it
is possible to be "close" to a target query vertex and "far" from all of its
neighbors, making pruning impossible. To overcome this, ANNS indices convert
MIPS into a Euclidean space by adding one extra dimension. This allows for
graph construction but comes at the cost of one extra dimension.

Both the \sysname and \sysnameRQ implementations suffer from reduced
performance on this dataset, as the conversion from MIPS to Euclidean distance
reduces the accuracy compared to systems like \cagra that can construct the
graph using the original MIPS distance calculation. \sysname has the highest
recall/throughput trade-off on this dataset until .9 recall, at which point
\cagra shows the highest recall/throughput trade-off. Due to the low dimensions
(200), \sysname exact shows higher performance than \sysnameRQ, with \sysnameRQ
showing reduced recall due to the MIPS distance calculation.


\begin{table}[h]
  \centering
  \begin{tabular}{lrrr}
  \toprule
  Dataset        & \texttt{mark\_deleted} & \texttt{consolidate}    & \texttt{compact}     \\
                 & (vert./s)              & (tombstones/s)          & (live-vert./s)       \\
  \midrule
  \gist 1M         & $1.26 \times 10^{8}$   & $2.36 \times 10^{4}$    & $2.37 \times 10^{7}$ \\
  \bigann 10M      & $2.76 \times 10^{8}$   & $4.71 \times 10^{4}$    & $1.69 \times 10^{8}$ \\
  \deep 10M        & $3.01 \times 10^{8}$   & $7.22 \times 10^{4}$    & $1.04 \times 10^{8}$ \\
  \openai     & $1.36 \times 10^{8}$   & $5.29 \times 10^{3}$    & $1.60 \times 10^{7}$ \\
  \textToImage 10M  & $2.82 \times 10^{8}$   & $2.39 \times 10^{5}$    & $7.71 \times 10^{7}$ \\
  \bottomrule
  \end{tabular}
  \caption{Throughput of the three deletion phases on A100 when deleting $10\%$ of the dataset.}
  \label{tab:deletion_throughput_a100}
  \vspace{-2em}
\end{table}

\begin{table}[h]
  \centering
  \begin{tabular}{lrrrr}
  \toprule
                & \multicolumn{2}{c}{Baseline} & \multicolumn{2}{c}{Post-consolidation} \\
  \cmidrule(lr){2-3} \cmidrule(lr){4-5}
  Dataset       & Recall@10 & QPS       & Recall@10 & QPS       \\
  \midrule
  \bigann 10M     & 0.986 & 574{,}990 & 0.988 & 563{,}998 \\
  \gist 1M        & 0.922 & 54{,}471  & 0.925 & 42{,}539  \\
  \bottomrule
  \end{tabular}
  \caption{Live recall@10 and query throughput before and after deletion+consolidation on A100 for small and large vectors ($10\%$ of dataset deleted, beam width 128).}
  \label{tab:deletion_recall_throughput}
  \vspace{-2em}
  \end{table}

\subsection{Deletion}
\label{subsec::deletion}

\RThreeInline{W2}{
Deletion in \sysname is split into three phases: mark, consolidate, and
compact. The performance of each of these phases of deletion can be seen in
~\Cref{tab:deletion_throughput_a100}. While consolidation is expensive due to
the need to search for every node that points to a deleted vertex, it is only
needed periodically; our empirical results show that up to 10\% of the dataset
can be atomically marked as deleted without affecting output quality or
throughput.}

\RThreeInline{W2}{
Query throughput before and after deletion can be seen
in~\Cref{tab:deletion_recall_throughput}. For both small and large vectors, the
recall is unaffected by deletion. For BigANN, the throughput is largely
unaffected by deletion, and graph degree is mostly stable, going from an
average degree of 44.2 to 45.0. For Gist, the throughput drops as the graph is
more densely connected after deletion: the deletion policy causes many new
connections to be added, resulting in average graph density growing from 24.4
to 30.6 after consolidation.}

\begin{figure}[t]
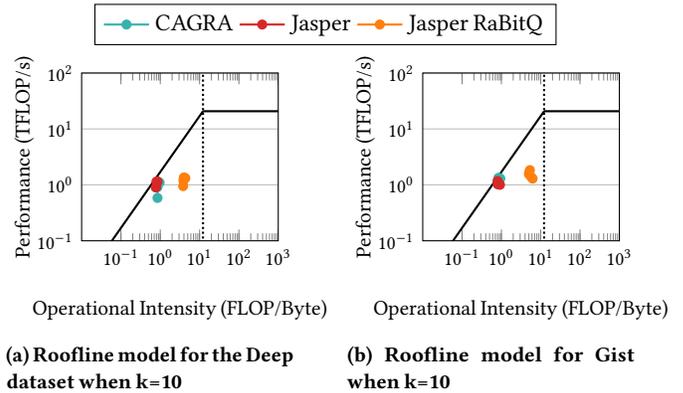

    \centering
    \ref{rooflineLegend}
    \input{figures/roofline/deep_roofline}\hfill
    \input{figures/roofline/gist_roofline}
    \caption{Roofline analysis of the \deep and \gist search kernels for \sysname, \sysnameRQ, and \cagra. Y-axis is on log-scale, higher is better.}
    \label{fig::roofline}
    \vspace{-1.5em}
\end{figure}

\subsection{Performance analysis} \label{eval:roofline}

\para{Roofline analysis} We characterize \sysname's performance on both
low-dimensional (\deep) and high-dimensional (\gist) vectors using the roofline
model~\cite{Williams2009}. \Cref{fig::roofline} compares the arithmetic
intensity and achieved throughput of \sysname, \sysnameRQ, and \cagra.

Across both datasets, the roofline results indicate that \sysname operates
predominantly in a memory-bound regime, with operational intensities are
clustered tightly around 0.7–0.95 FLOP/byte. On DEEP, \sysname achieves peak
throughputs around 0.89-1.14 TFLOP/s at these low intensities, closely tracking
the sloped bandwidth roof. This behavior is consistent with ANNS workloads
dominated by irregular memory accesses during graph traversal, where additional
compute cannot be fully exploited without increasing data reuse. The comparable
intensity–performance envelope between \sysname and \cagra on \deep suggests
that both systems are similarly constrained by memory bandwidth, with
performance differences largely attributable to constant-factor effects such as
traversal efficiency and kernel scheduling rather than fundamental compute
utilization.

In contrast, \sysnameRQ significantly increases operational intensity, where
intensities rise to 5.0–6.2 FLOP/byte and achieved performance reaches up to
~1.83 TFLOP/s. This rightward shift on the roofline indicates substantially
improved arithmetic reuse, allowing \sysnameRQ to move closer to the compute
roof and escape the strict bandwidth-bound regime observed in \sysname.
Overall, these results show that \sysnameRQ’s reordering and query-time
computation strategies improve performance by reducing memory traffic. 




\begin{figure}[t]
    \centering
    \ref{ablationLegend}
    \vspace{.2em}

    \input{figures/ablation/bigann_subfig}%
    \hfill
    \input{figures/ablation/gist_subfig}

    \caption{Ablation results for BigANN and Gist datasets.}
    \label{fig::ablation}
    \vspace{-1em}
\end{figure}

\para{Ablation study}
\RFourInline{W1}{}
\RThreeInline{D3}{
\Cref{fig::ablation} shows the ablation results. We start at a fully optimized kernel and remove
optimizations. We start at a full implementation with \RQ and ablate away
\RQ, chunked loads, optimized tile size, the use of a visited bit instead of
a hash table, and optimized block size in that order.
}

\RThreeInline{D3}{
For low-dimensional vectors such as \bigann, the ablation optimizations besides
\RQ are purely optimal, with performance increasing at all beam widths for
all optimizations. For \gist, All optimizations are optimal expect for the use
of visited bits, where the use of a hash table is better for all beam widths
when ablated. This speedup doesn't hold when all optimizations are applied due
to the optimized tile size: with more threads per tile, the amount of time
taken for distance calculations reduces, which lowers the penalty of
recalculating distances when using the visited set.
}

\section{Conclusion} \label{sec:conclusion}

We showed that GPU-native approximate nearest neighbor search (ANNS) can
substantially outperform CPU-based approaches while enabling high-throughput
retrieval and efficient co-location with downstream machine learning workloads.
Our study further reveals that modern GPUs expose new bottlenecks: existing GPU
ANNS systems often underutilize hardware due to synchronization overheads and
excessive shared memory usage. By simplifying search kernels and optimizing
memory access, \sysname achieves up to \timesFasterThanCagra$\times$ higher
query throughput than CAGRA while supporting streaming updates, delivers
\lowerBangSpeedup--\higherBangSpeedup$\times$ higher throughput than BANG, and
constructs indexes \avgConstruct faster than CAGRA on average. Our optimized
search kernel sustains 70--80\% of peak theoretical throughput.

Roofline analysis shows that state-of-the-art GPU ANNS kernels are already
close to the memory-bandwidth limit, suggesting that future gains will
primarily come from reducing data movement, for example through
quantization-aware designs such as RaBitQ. Scaling beyond GPU memory remains an
open challenge: even the largest GPUs can store graphs for only about 200
million vectors, requiring multi-GPU sharding or host-pinned memory for larger
datasets. Developing efficient architectures for this setting is an important
direction for future work.

\balance

\iftruncate
\section{Acknowledgments}
This research is funded in part by NSF grants OAC 2339521, 2517201, and 2513656. Additionally, we
would like to thank Harsha Vardhan Simhadri and Magdalen Dobson Manohar for
insightful discussions early in the project.
\fi

\clearpage
\bibliographystyle{ACM-Reference-Format}
\bibliography{ann_bib}

\ifappendix

\clearpage
\appendix

\noindent{\huge\bfseries Appendix}\par\vspace{0.5em}

\begin{algorithm}
\DontPrintSemicolon
\caption{RobustPrune (DiskANN)}
\label{alg:robustprune}
\KwIn{Node $p$, candidate neighbors $N(p)$, degree cap $R$, pruning factor $\alpha$, distance $d$}
\KwOut{Pruned neighbor set $N'(p)$}
Sort $N(p)$ by $d(p, \cdot)$ (closest first)\;
$N'(p) \gets \emptyset$\ (initialize as empty);

\While{$|N'(p)| < R$ and $N(p)$ not empty}{
    $p^* \gets$ extract closest from $N(p)$\;
    Add $p^*$ to $N'(p)$\;
    \For{each $p' \in N(p)$}{
        \If{$\alpha \cdot d(p^*, p') \leq  d(p, p')$}{
            Discard $p'$ from $N(p)$\;
        }
        \Else{
            Keep $p'$ in $N(p)$\;
        }
    }
}
\Return $N'(p)$
\end{algorithm}

\begin{algorithm}
\DontPrintSemicolon
\caption{Batch Insertion (ParlayANN)}
\label{alg:parlayann}
\KwIn{Batch of new points $B$, current graph $G$, beam width $K$, degree cap $R$}
\KwOut{Graph $G$ updated with $B$}

\textbf{Step 1: Local candidate generation}\;
\ForPar{each $x \in B$}{
    $(F, V) \gets \text{BeamSearch}(G, x, K, G.medoid)$\;
    $E_x \gets \{(x, v): v \in V\}$ \tcp*{add edges for new nodes}
}

\textbf{Step 2: Global edge collection}\;
$E \gets \bigcup_{x \in B} E_x$ \tcp*{concatenate edges}

\textbf{Step 3: Semisort and parallel pruning}\;
Group $E$ by endpoint using \textbf{Semisort}\;
\ForPar{each vertex $u$ touched by $E$}{
    $N'(u) \gets \text{RobustPrune}(u, N(u) \cup E[u], R)$\;
    Update adjacency list of $u$ with $N'(u)$\;
}

\Return $G$
\end{algorithm}

\section{Algorithm code}
\label{app:algorithms}

We include pseudocode for both the \prune procedure used in all DiskANN systems (\Cref{alg:robustprune}) and the bulk-synchronous batch insertion procudeure from \parlayann that is used in \sysname (\Cref{alg:parlayann}).

\section{Search Kernel Optimizations}
\label{app:kernel-opt}

We describe the low-level kernel optimizations used to adapt greedy beam
search to the GPU execution model. We focus on reducing redundant arithmetic
intensity, improving memory efficiency, and maximizing parallel utilization
within each thread block, enabling distance evaluation and candidate processing
to approach the hardware limits of modern GPUs.

\para{Optimizing distance calculations}
We observe that the square root operation accounts for over 40\% of Euclidean
distance computation time. Since square root is monotonic over the non-negative
reals, we safely elide it and compare squared distances instead, preserving
correctness while significantly reducing compute overhead.

\para{Optimizing beam search through kernel fusion}
We further reduce overhead by fusing distance computation, frontier sorting,
and neighbor expansion into a single kernel executed by one thread block per
query (illustrated in~\Cref{fig:query-gpu}). Kernel fusion
eliminates repeated launch overhead and allows all intermediate data to remain
in shared memory, improving both latency and bandwidth utilization, similar to
the approach used in \cagra~\cite{ootomo2024}.

\section{Additional Hardware Results}
\label{app:hardware}

\subsection{Consumer GPU: NVIDIA RTX 4080 Super (10M datasets)}
\label{app:4080}

\begin{figure*}[!htbp]
\centering
\ref{queryLegend4080}

\setlength{\tabcolsep}{0pt}

\begin{tabular}{c c}
\makebox[10pt][c]{\rotatebox[origin=c]{90}{\large Throughput in queries/secs (Log-Scale)}} &
\begin{tabular}{c c c c}

\input{figures/4080-bigann-k1} &
\input{figures/4080-bigann-k10} &
\input{figures/4080-bigann-k50} &
\input{figures/4080-bigann-k100} \\

\input{figures/4080-deep-k1} &
\input{figures/4080-deep-k10} &
\input{figures/4080-deep-k50} &
\input{figures/4080-deep-k100}

\end{tabular}
\end{tabular}

\vspace{0.5em}
\makebox[\textwidth][c]{\large Recall (\%)}

\caption{Query recall/throughput on the NVIDIA RTX 4080 Super (16~GB GDDR6X)
for the two datasets that fit in device memory at 10M points. Y-axis is on
log-scale; higher and more to the right is better.}
\label{fig::4080_query}
\vspace{-0.5em}
\end{figure*}

\RThreeInline{D2}{To evaluate \sysname on a lower-cost consumer GPU, we
repeat the query benchmark on an NVIDIA RTX 4080 Super (16~GB GDDR6X,
PCIe~4.0, 736~GB/s peak memory bandwidth). The two datasets that fit
within 16~GB are \bigann 10M and \deep 10M; larger datasets and the
high-dimensional vector workloads exceed device memory at this scale.
\Cref{fig::4080_query} reports the recall--throughput curves for \sysname,
\sysnameRQ, and \cagra across $k\!\in\!\{1,10,50,100\}$. We restrict the
comparison on this platform to the GPU-native graph indices that fit
inside the 16~GB device budget; \bang, \ganns, and \parlayann were not
run on the 4080.}

\RThreeInline{D2}{Two patterns from the A100 results recur on the 4080.
First, on the memory-bound \bigann workload \sysname is Pareto-optimal
over the recall range it reaches. Absolute throughput is approximately
3$\times$ lower than on the A100, which is consistent with the 4080
Super's reduced memory bandwidth relative to the A100. Second, on \deep
the crossover regime in which \cagra reaches higher recall than
\sysnameRQ on 50@50 and 100@100 queries persists on the 4080, suggesting
that this regime reflects algorithmic differences rather than an
A100-specific artifact.}

\RThreeInline{D2}{We report results using the block size and tile width
chosen for the A100; we did not separately re-tune these parameters on
the 4080. The configurations remained effective in this consumer-GPU
setting, although a dedicated sweep on the 4080 may reveal further
opportunities.}

\subsection{Data Center GPU: NVIDIA H200 (200M datasets)}
\label{app:h200}

\begin{figure*}[!htbp]
\centering
\ref{queryLegendH200}

\setlength{\tabcolsep}{0pt}

\begin{tabular}{c c}
\makebox[10pt][c]{\rotatebox[origin=c]{90}{\large Throughput in queries/secs (Log-Scale)}} &
\begin{tabular}{c c c c}

\input{figures/h200-deep-k1} &
\input{figures/h200-deep-k10} &
\input{figures/h200-deep-k50} &
\input{figures/h200-deep-k100}

\end{tabular}
\end{tabular}

\vspace{0.5em}
\makebox[\textwidth][c]{\large Recall (\%)}

\caption{Query recall/throughput on the NVIDIA H200 (140~GB HBM3E) for
\deep at 200M points. Y-axis is on log-scale; higher and more to the right is
better. \cagra was forced onto its IVF construction path because NN-Descent
exceeded available device memory at this scale (see~\Cref{app:memory}).}
\label{fig::h200_query}
\vspace{-0.5em}
\end{figure*}

\RFourInline{W3}{To evaluate \sysname at a scale that exceeds the A100's
device memory, we repeat the query benchmark on an NVIDIA H200 (140~GB
HBM3E, 4.8~TB/s peak memory bandwidth) using \deep at 200M points.
\Cref{fig::h200_query} reports the recall--throughput curves.}

\RFourInline{W3}{In our runs, \cagra's NN-Descent construction did not
complete within the H200's 140~GB device budget at 200M points and the
system fell back to its IVF construction path; the breakdown of working
sets is reported in~\Cref{app:memory}. The \cagra curves in
\Cref{fig::h200_query} therefore correspond to an IVF-built graph rather
than the NN-Descent graph used in the main paper's 100M comparisons.
\sysname's batch-parallel Vamana construction stays within device memory
and produces a graph whose recall--throughput envelope at 200M tracks the
envelope observed at 100M.}

\section{Construction Memory Footprint}
\label{app:memory}

\RTwoInline{W2}{In this section, we evaluates the GPU memory usage between
\sysname and \cagra's NN-descent algorithm during index construction, 
shown in \cref{tab:memory}.}

\begin{table}[ht]
\centering
\begin{tabular}{lrr}
\hline
 & \textbf{\sysname} & \textbf{\cagra} \\
\hline
\textbf{BigANN}               & 51.20 GB & 145.29 GB \\
\textbf{Deep}                 & 44.80 GB  & 139.33 GB \\
\textbf{Gist}                 & 2.18 GB  & 3.00 GB    \\
\textbf{OpenAI-Arxiv}         & 7.65 GB & 9.37 GB   \\
\textbf{Yandex Text-to-Image} & 6.56 GB  & 15.87 GB  \\
\hline
\end{tabular}
\caption{Total construction memory usage comparison}
\label{tab:memory}
\end{table}

\RTwoInline{W2}{For \sysname, the memory usage is calculated by adding the raw
vector size and the graph size. \sysname and \cagra store every float vector as \texttt{\_\_half} (FP16).
Let $N$ be the number of vectors,
$D$ be the vector dimension, $K$ be the graph degree (64), $s$ be the index
type size, and $v$ be the vector type size (2 bytes for FP16). 
\sysname uses $N(Dv+Ks)$ bytes to store the vectors and the graph. 
Since our construction batch size is
an adjustable parameter, we can use whatever memory is left on the current GPU.}

\RTwoInline{W2}{For \cagra memory estimation, we follow the allocation heuristics
on \href{https://github.com/rapidsai/cuvs/blob/main/cpp/src/neighbors/nn_descent.cu}{GitHub}.
Let $N$ be the number of vectors, $D$ be the vector dimension, $K$ be the graph degree (64),
and $s$ be the index type size (4-byte \texttt{uint32\_t}). NN-descent
requires $2ND+4NK(s+1)+24N$ bytes. The leading $2ND$ term reflects \cagra
storing all datasets as \texttt{\_\_half} (FP16), the same as \sysname.}


\section{Microbenchmarks}
\label{sec::microbenchmark}

We analyze key design decisions in \sysname through a series of microbenchmarks.

We present microbenchmarks on block size (\Cref{app:block-sweep}), tile size (\Cref{app:tile-sweep}), coalesced
versus tiled loading (\Cref{app:load-microbench}), and quantization methods
(\Cref{app:quant-microbench}).

\input{figures/pq_vs_rq}
\section{Quantization Microbenchmark}
\label{app:quant-microbench}

We compare quantization methods in~\Cref{fig:quantization}, measuring query
throughput for \sysname (exact and RaBitQ) against \cagra (exact and PQ) and
\parlayann (exact and 8-bit scalar quantization). We evaluate \cagra
s PQ using their default settings. For
\parlayann's SQ, we uses 8 quantized bits per dimension.

\cagra's exact variant is optimized for high-dimensional vectors and slightly
outperforms \sysname's exact computation. However, \sysnameRQ outperforms all
\cagra variants, including exact. \cagra's PQ achieves throughput similar to
its exact version despite reducing memory footprint, the scattered lookups
required for PQ distance computation negate any bandwidth savings. These
results confirm that RaBitQ is more effective than PQ for GPU workloads.
\parlayann also benefits from reduced memory footprint from quantitation on
CPU, showing 3 to 4 times speed up for the same recall compare to \parlayann
exact version.

\section{Block-Size Microbenchmark}
\label{app:block-sweep}

\begin{figure}[t]
    \centering
    \ref{blockSizeLegend}
    \input{figures/block_size_bigann}\hfill
    \input{figures/block_size_gist}
    \caption{Varying Block size for \bigann and \gist. Throughput is in millions of queries per second, higher is better.}
    \label{fig::block_size}
\end{figure}
\input{figures/block_size_appendix}

Block size controls the trade-off between parallelism within a query and the
number of concurrent queries per SM. Larger blocks assign more threads to each
query; smaller blocks allow more queries to execute simultaneously, increasing
memory-level parallelism. \RThreeInline{D3}{\Cref{fig::block_size_bigann}
and~\Cref{fig::block_size_gist} report results at the optimal beam width for
each dataset; the two highlighted configurations are the optimal block size for
each dimensionality regime. For completeness, \Cref{fig::block_size_appendix}
reports the full sweep across 32, 64, 128, 256, 512, and 1024 threads per block
on \bigann, \deep, and \gist at $k$@$k\!=\!1$@$1$.}

For datasets with high-dimensional vectors (\gist, \openai), reducing block
size improves performance until reaching 128 threads per block. Below this
point, performance drops because distance computation becomes the
bottleneck—there are too few threads to efficiently process each
high-dimensional vector.
For datasets with low-dimensional vectors (\bigann, \deep), distance
computation is not the bottleneck. Smaller block sizes consistently improve
performance by enabling greater memory parallelism. The optimal configuration
uses 32 threads per block (a single warp per query), allowing 32 concurrent
queries per SM and maximizing memory throughput.

\RThreeInline{D3}{The sweep is consistent with the pattern
described in the design section. On the low-dimensional workloads
(\bigann, \deep) throughput is maximized at 32 threads per block:
shrinking the block raises the number of concurrent queries per SM and thus
the corresponding memory-level parallelism. Moving from 32 to 1024 threads per block reduces
throughput at fixed recall on \bigann by roughly an order of magnitude,
which we attribute to the proportional reduction in concurrent queries
per SM. On the high-dimensional \gist workload the trend reverses:
distance computation dominates and benefits from larger per-query
parallelism; the observed optimum sits at 128--256 threads per block. No
single fixed block size is optimal across these datasets, motivating the
dataset-adaptive choice described in~\Cref{subsec:gpu-optimizations}.}

\section{Tile Size Microbenchmark}
\label{app:tile-sweep}

\begin{figure}[t]
    \centering
    \ref{tileSizeLegend}
    \input{figures/tile_size_bigann}\hfill
    \input{figures/tile_size_gist}
    \caption{Varying Tile size for \bigann and \gist. Throughput is in millions of queries per second, higher is better.}
    \label{fig::tile_size}
\end{figure}

The tile microbenchmark measures the performance of
different tile sizes on exact vector query when querying \bigann 1@1 and \gist
1@1. The results for this benchmark can be seen in \Cref{fig::tile_size}. For
large tile sizes, compute is maximized (32 threads running every cycle), but
the amount of memory that can be loaded is limited, as the entire tile is
working on one vector. Halving the tile size has the following effects: it
doubles the amount of distance calculations that can be done simultaneously,
halves the number of workers per distance calculation, and halves the frequency
at which each tile is scheduled to work. This creates a quadratic tradeoff
curve - doubling the number of memory requests issued by the warp reduces the
compute available per team by $4\times$. For this reason, a tile size of one is the
slowest tile size even though it offers the most memory parallelism: individual
workers are scheduled infrequently and are too slow to efficiently calculate
disrtance. We find that the best tile sizes for both large and small vectors
\gist is 8 as it strikes the most performant balance between memory and compute
parallelism.

\section{Load Microbenchmark}
\label{app:load-microbench}


    
    
    


\setlength{\tabcolsep}{3.5pt}
\begin{table}
\begin{center}
\begin{tabular}{|c|c c c|c c c|} 
 \toprule
  & \multicolumn{3}{c|}{Beam width=1} & \multicolumn{3}{c|}{Beam width=256}\\
  & \bigann&\deep&\gist&\bigann&\deep&\gist\\  
 \midrule
 Tiled &8.398& 6.701 & .690 &.095 &.089& .024\\
 Chunked & 9.611 & 7.622 & .769 & .095 & .089 & .024\\
 \bottomrule
\end{tabular}
\end{center}
\caption{Difference in tiled and chunked load strategies. Table results are throughput in millions of queries per second. For smaller beam widths, the chunked strategy provides lower latency as all load requests are issued simultaneously. For higher beam widths, total memory throughput dominates, resulting in identical performance.}
\label{tab::load_comparison}
\end{table}

\para{Load microbenchmark} The load microbenchmark measures performance of our
coalesced load implementation against a traditional tiled approach. In this
microbenchmark we run both approaches and query \bigann, \deep, and \gist with
beam widths 1 and 256, the smallest and largest beam widths supported. Float vectors are stored as full 32-bit floats and distances are calculated with 32-bit floats. The
results for this benchmark are in~\Cref{tab::load_comparison}. For small beam
widths, we see that the performance of the coalesced loads is consistently
higher than tiled loads. This is due to the reduced latency from issuing load
instructions simultaneously. For low beam widths, using coalesced loads
provides a 14\% performance increase as loads are sent and thus return faster,
reducing the latency of the query. For larger beam widths, latency hiding means
that total memory bandwidth dominates performance, and the coalesced loads show
no measurable change in performance.

\para{Tile microbenchmark}
We compare our coalesced load implementation against traditional tiled loading
on \bigann, \deep, and \gist using beam widths of 1 and 256
(\Cref{tab::load_comparison}). At small beam widths, coalesced loads provide a
14\% throughput improvement by issuing all load instructions simultaneously,
reducing latency. At large beam widths, sufficient in-flight operations hide
memory latency, so both approaches achieve similar throughput.


\fi

\end{document}